\documentclass[aps,prb,amsmath,amssymb,footinbib,showpacs,twocolumn,superscriptaddress]{revtex4-2}
\usepackage{amsmath}
\usepackage{amssymb}
\usepackage{amsthm}
\usepackage{setspace}
\usepackage{graphicx}
\usepackage{braket}
\usepackage{mathrsfs}
\usepackage{float}
\usepackage[colorlinks = true,linkcolor = blue,urlcolor  = blue,citecolor = blue,anchorcolor = blue]{hyperref}
\usepackage[utf8]{inputenc}
\usepackage[english]{babel}
\usepackage{bm}

\begin{document}

\title{Longitudinal  magnetoconductance and the planar Hall conductance in inhomogeneous Weyl semimetals}

\author{Azaz Ahmad}
\affiliation{School of Physical Sciences, Indian Institute of Technology Mandi, Mandi 175005, India}
\author{Karthik V. Raman}
\affiliation{Tata Institute of Fundamental Research, Hyderabad, Telangana 500046, India}
\author{Sumanta Tewari}
\affiliation{Department of Physics and Astronomy, Clemson University, Clemson, South Carolina 29634, USA}
\author{G. Sharma}
\affiliation{School of Physical Sciences, Indian Institute of Technology Mandi, Mandi 175005, India}

\begin{abstract}
Elastic deformations (strain) couple to the electronic degrees of freedom in Weyl semimetals as an axial magnetic field (chiral gauge field), which in turn affects their impurity dominated diffusive transport. Here we study the longitudinal magnetoconductance (LMC) in the presence of strain, Weyl cone tilt, and finite intervalley scattering, taking into account the momentum dependence of the scattering processes (both internode and intranode), as well as charge conservation. We show that strain induced chiral gauge field results in `strong sign-reversal' of the LMC, which is characterized by the reversal of orientation of the magnetoconductance parabola with respect to the magnetic field.  On the other hand, external magnetic field results in `strong sign-reversal', only for sufficiently strong intervalley scattering. When both external and chiral gauge fields are present, we observe both strong and weak sign-reversal, where in the case of weak sign-reversal, the rise and fall of magnetoconductivity depends on the direction of the magnetic field and/or the chiral gauge field, and is not correlated with the orientation of the LMC parabola.
The combination of the two fields is shown to generate striking features in the LMC phase diagram as a function of various parameters such as tilt, strain, and intervalley scattering. We also study the effect of strain induced chiral gauge field on the planar Hall conductance and highlight its distinct features that can be probed experimentally.  
\end{abstract}

\maketitle

\section{Introduction}
Fermions and the atomic lattice form the building blocks of condensed matter. While each of them are fundamentally different from the other, the interplay between the two leads to remarkable effects. In recent works, massless Dirac fermions, which have resurged in condensed matter, have been shown to couple to the elastic deformations of the lattice (strain) as an axial magnetic field (also known as chiral gauge field). Prominent examples where such fields can be realized include graphene~\cite{jackiw2007chiral,vozmediano2010gauge,guinea2010energy} and three-dimensional Weyl semimetals~\cite{cortijo2015elastic,pikulin2016chiral,grushin2016inhomogeneous}. For instance, in graphene, the generated field can be even as large as 300T, as observed via spectroscopic measurement of the Landau levels~\cite{levy2010strain}. A measurement of strain induced chiral magnetic field as well as its implications on electron transport in three-dimensional Weyl and Dirac semimetals materials is of high interest to the condensed matter community.

The reason why Weyl and Dirac semimetals also have been fascinating is due to some intriguing properties that are absent in conventional metals. Some examples include the anomalous Hall~\cite{yang2011quantum,burkov2014anomalous} and Nernst~\cite{sharma2016nernst,sharma2017nernst,liang2017anomalous} effects, open Fermi arcs~\cite{wan2011topological}, planar Hall and Nernst effects~\cite{nandy2017chiral,sharma2019transverse}, and the manifestation of chiral or Adler-Bell-Jackiw anomaly~\cite{adler1969axial,nielsen1981no,nielsen1983adler,bell1969pcac,aji2012adler,zyuzin2012weyl,zyuzin2012weyl,son2012berry,goswami2015optical, fukushima2008chiral,goswami2013axionic}. The origin of each of these effects can be traced down to the non-trivial topology of the Bloch bands. Specifically, the low-energy bandstructure of Weyl nodes comprise of pairs of non-degenerate massless Dirac cones that are topologically protected by the chirality quantum number (also known as the Chern number). Without any coupling to an external gauge field, the charge of a given chirality remain conserved. The conservation law is however broken when Weyl fermions are coupled to background gauge fields such as electric or magnetic fields~\cite{adler1969axial,nielsen1981no,nielsen1983adler}. This breakdown of conservation laws is known as `chiral anomaly', rooting its name from the particle physics literature. The verification of chiral anomaly in  Weyl semimetals is a very active area of investigation in condensed matter physics.

In a minimal model of Weyl semimetal, Weyl nodes must be separated in momentum space by a vector $\mathbf{b}$ to ensure topological protection. Alternatively, the vector $\mathbf{b}$ can also be interpreted as an axial gauge field since it couples with an opposite sign to Weyl nodes of opposite chirality~\cite{goswami2013axionic,volovik1999induced,liu2013chiral,grushin2012consequences,zyuzin2012topological}. Thus spatial variation of $\mathbf{b}$ generates an axial magnetic field $\mathbf{B}_5=\nabla\times \mathbf{b}$, which also couples oppositely to Weyl nodes of opposite chirality. An effective $\mathbf{B}_5$ field can emerge from an inhomogeneous strain profile in Weyl semimetals. In the presence of an effective chiral gauge field $\mathbf{B}_5$, the effective magnetic field experienced by Weyl fermions at a given node of chirality $\chi$ is $\mathbf{B}\longrightarrow\mathbf{B}+\chi\mathbf{B}_5$, where $\mathbf{B}$ is the external magnetic field. Therefore, the conservation laws are also modified accordingly  in the presence of the $\mathbf{B}_5$ field. Recent works have pointed out that even in the absence of an external magnetic field, the chiral gauge field influences the diffusive electron transport in Weyl semimetals by modifying its longitudinal magnetoconductance (LMC)~\cite{grushin2016inhomogeneous} as well as the planar Hall conductance (PHC)~\cite{ghosh2020chirality}. Although true in spirit, the drawback of these works is that they ignore the momentum dependence of scattering when the Weyl fermions scatter within a node (known as intranode scattering or intravalley scattering) conserving both the total charge and chiral charge, and also when they scatter to the other node (internode/intervalley scattering), in which case they conserve only the total charge. Moreover, intervalley scattering, which is the essence of `true chiral anomaly', has been neglected in Ref.~\cite{ghosh2020chirality}. In a recent work~\cite{sharma2022revisiting}, some of the co-authors of this work have pointed out that momentum dependence of scattering as well as charge conservation constraint can lead to drastic differences in the qualitative conclusions. It is therefore of immense importance to correctly treat the effect of strain induced gauge field on electron transport in Weyl semimetals, which is the focus of this work.

\begin{figure}
    \centering
    \includegraphics[width=\columnwidth]{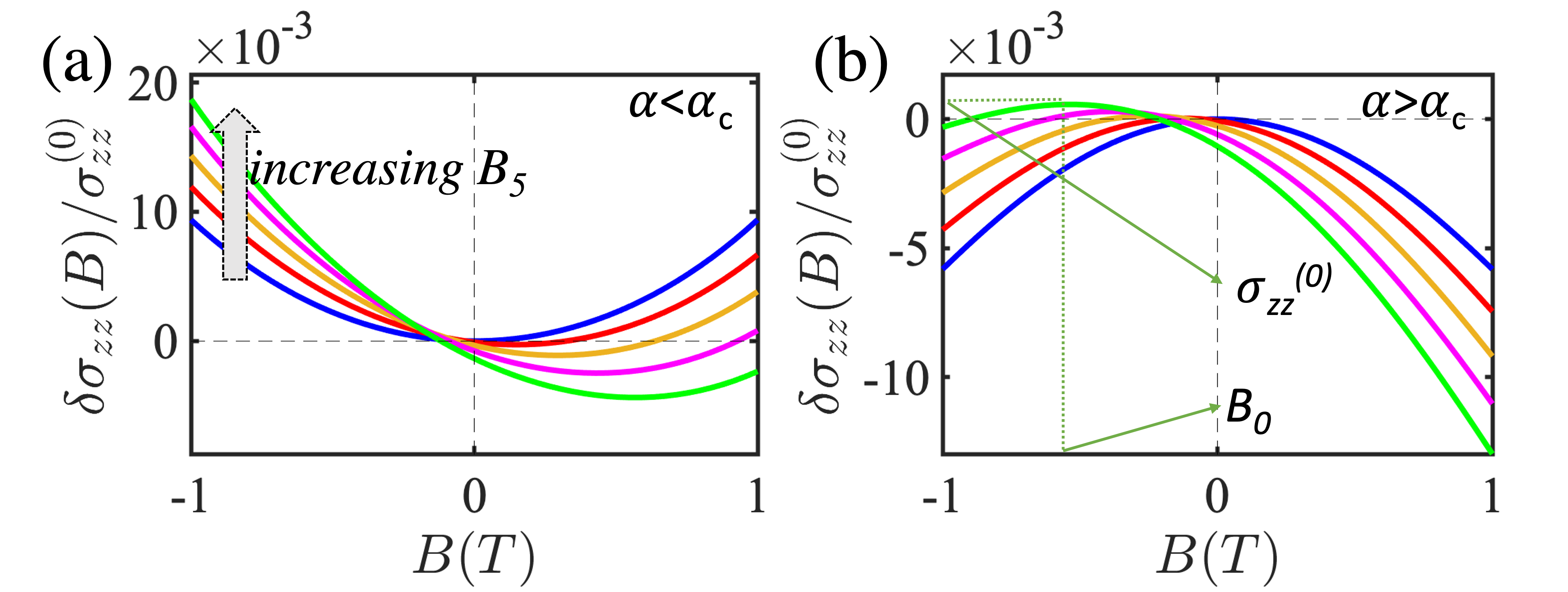}
    \caption{Change in LMC ($\delta\sigma_{zz}(B)$) with respect to the magnetic field for a minimal model of untilted TR broken WSM (Eq.~\ref{Eq:sigz1}). (a)  Weak intervalley scattering ($\alpha<\alpha_c$), and (b) strong (and weak) intervalley scattering ($\alpha>\alpha_c$). As we move from  blue to the green curve in both the plots (in the direction of the arrow), we increase $B_5$ from zero to 0.2T. The $B_5$-field is held parallel to the external magnetic field. The vertex $B_0$ and the corresponding $\sigma_{zz}^{(0)}$ is marked for the green curve in plot (b).}
    \label{fig:sigma001}
\end{figure}

In this work we critically examine the effect of strain induced chiral gauge field via the Boltzmann formalism (thus limiting ourselves to only weak perturbative fields) on two linear response quantities: the longitudinal magnetoconductance, and the planar Hall conductance. We study these effects in both time-reversal breaking WSM (with and without tilt) as well as inversion asymmetric Weyl semimetals. Earlier it was believed that positive longitudinal magnetoconductivity must manifest from chiral anomaly at least in the limit of weak external magnetic field, but this claim was corrected later on when sufficiently strong intervalley scattering was shown to switch the sign of longitudinal magnetoconductivity even in the weak-$\mathbf{B}$ limit~\cite{knoll2020negative}.  Typically, by positive (negative) longitudinal magnetoconductance we mean that $(\sigma(|\mathbf{B}|) - \sigma(\mathbf{B}=0)) >(<)\; 0$, i.e., the field dependent conductivity is greater (smaller) than the zero-field conductivity. 
Here we show that the presence of $\mathbf{B}_5$ field can also reverse the sign of LMC, but along a particular direction of the magnetic field (see Fig.~\ref{fig:sigma001}). This leads to an interesting scenario of the LMC being positive along one direction of the magnetic field, and negative when the direction of the magnetic field is reversed. To counter this  ambiguity in the sign of LMC, we introduce the idea of weak and strong sign-reversal, which depends on the orientation and the vertex of the parabola of magnetoconductivity with respect to the magnetic field (Eq.~\ref{eq:sigma_2}). We show that in the presence of only strain induced chiral gauge field (and absence of external magnetic field), the system shows signatures of strong sign-reversal for all values of intervalley scattering. In the presence of only the external magnetic field (and absence of chiral gauge field), the system shows strong sign-reversal only at sufficiently large values of scattering. In the presence of both chiral gauge and externally applied magnetic field, signatures of both weak and strong sign-reversal are observed, and furthermore very interesting features emerge in the phase diagram of LMC as a function of various system parameters such as the intervalley scattering, tilt, and strain. We point out that whenever external magnetic field is absent, we discuss weak and strong-sign reversal in context of the LMC parabola with respect to the chiral gauge field $\mathbf{B}_5$. When the external magnetic field is present (in either presence or absence of the chiral gauge field), weak and strong-sign reversal in discussed in context of the LMC parabola with respect to the external magnetic field $\mathbf{B}$.
We also extend the idea of weak and strong sign-reversal to the planar Hall conductance as well, and study the effect of strain induced gauge field on the same. Along with other features, we also unravel a very interesting behavior in the planar Hall conductance due to an interplay between the chiral gauge field and the external magnetic field. Specifically we observe a region in the parameter space where the planar Hall conductivity increases in magnitude upon increasing the scattering strength, which is counter-intuitive.  
In Section II, we introduce the concept of weak and strong sign-reversal using a minimal model of a TR broken WSM. We also study the interplay of strain, tilt, and intervalley scattering on LMC and PHC. In Section III, we present the results for inversion asymmetric Weyl semimetals. We conclude in Sec IV. All the calculations are relegated to the Appendix. 

\section{Time-reversal broken Weyl semimetals}
Consider a minimal model of a time-reversal symmetry broken Weyl semimetal, i.e., two linearly dispersing non-degenerate Weyl cones separated in momentum space. We also assume that there is no tilting of the Weyl cones in any direction.  The low-energy Hamiltonian is given by 
\begin{align}
    H = \sum\limits_\chi \sum\limits_\mathbf{k} {\chi\hbar v_F \mathbf{k}\cdot\boldsymbol{\sigma}}
    \label{Eq:HWeyl1}
\end{align}
Here $\chi=\pm 1$ is the chirality of the node, $\mathbf{k}$ is the momentum, $v_F$ is the velocity parameter, and $\boldsymbol{\sigma}$ is the vector of Pauli spin matrices. Both intranode and internode scattering processes are allowed, and the dimensionless intervalley scattering strength is denoted by $\alpha$ (see Appendix A for all the calculations). To study transport, we perturb the system with weak electric field that is fixed along the $\hat{z}-$axis. On application of a magnetic field parallel to the electric field, the longitudinal magnetoconductivity obtained in the semiclassical limit is expressed as
\begin{align}
\sigma_{zz}(B)= \sigma_{zz}^{(2)}B^2 + \sigma_{zz}^{(0)},
\label{Eq:sigz1}
\end{align}
where $\sigma_{zz}^{(0)}$ is the conductivity in absence of any magnetic field, while $\sigma_{zz}^{(2)}$ is the quadratic coefficient of magnetic field dependence. In contrast to earlier anticipation that the quadratic coefficient $\sigma_{zz}^{(2)}$ is always positive, it was recently realized that the coefficient can become negative if the intervalley scattering is sufficiently strong~\cite{knoll2020negative}. In other words, large intervalley scattering results in negative longitudinal magnetoconductivity or reverses its sign. Specifically this occurs above a critical intervalley scattering strength $\alpha_c$. The sign of the parameter $\sigma_{zz}^{(2)}$ also correlates with increasing or decreasing longitudinal magnetoconductivity.  We can call this as the usual `sign-reversal' of LMC, which refers to the fact that $\sigma_{zz}(|B|)-\sigma_{zz}(B=0)$ continuously changes sign from positive to negative. 
\begin{figure}
    \centering
    \includegraphics[width=\columnwidth]{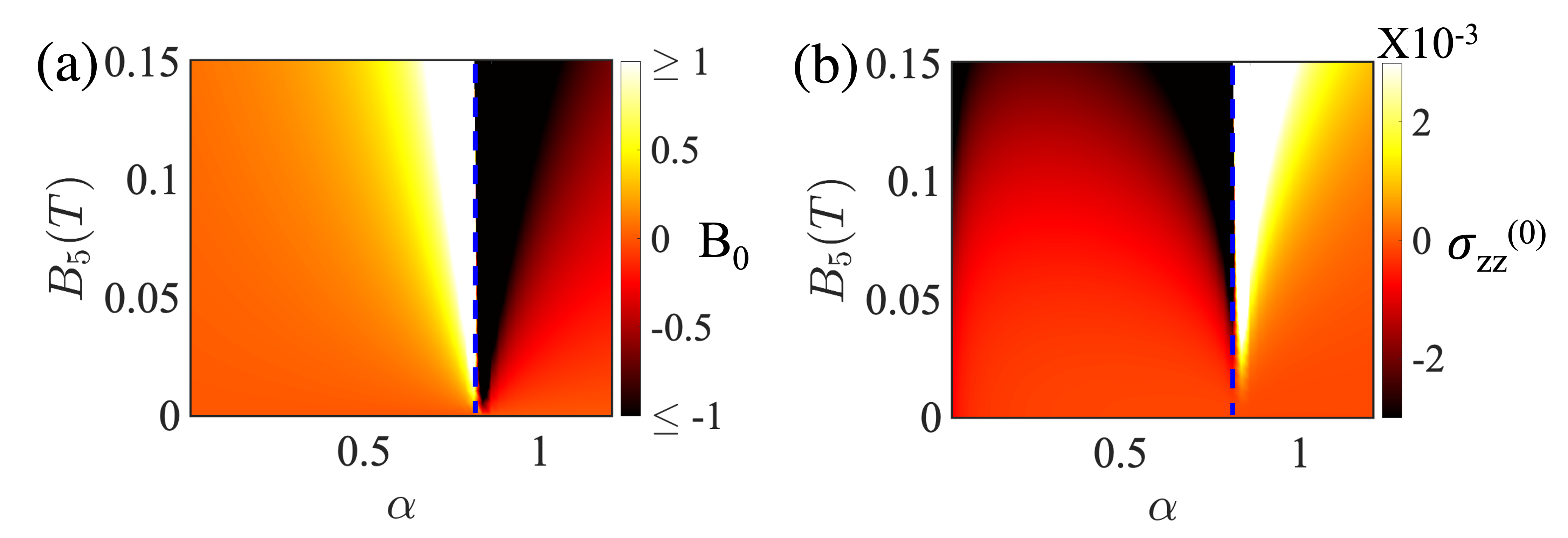}
    \caption{(a) The vertex of the parabola $B_0$, and (b) conductivity at $B_0$ for a minimal model of untilted TR broken WSM (Eq.~\ref{Eq:sigz1}). Around the blue dashed contour ($\alpha=\alpha_c$) we see `strong' sign-reversal. The parameters $B_0$ and $\sigma_{zz}^{(0)}$ show a striking change of sign as we move across the $\alpha_c$ contour. }
    \label{fig:sigma002}
\end{figure}
\begin{figure*}
    \centering
    \includegraphics[width=2\columnwidth]{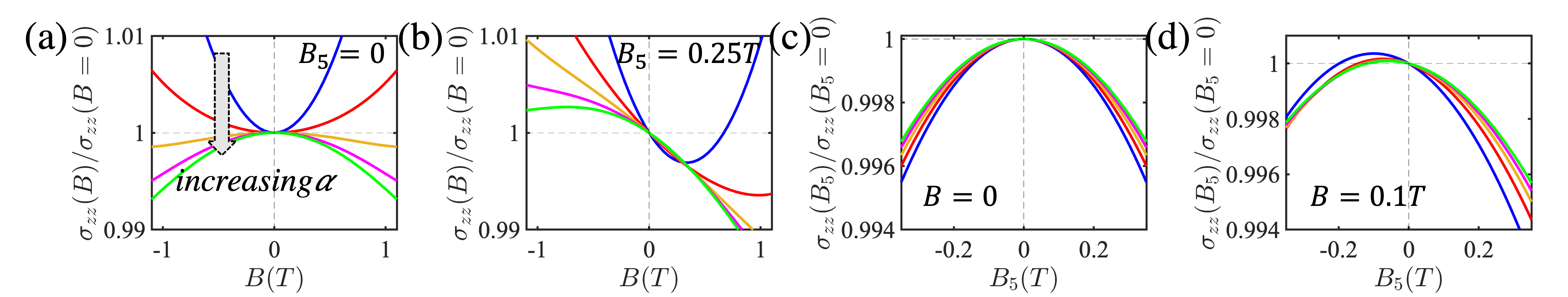}
    \caption{Longitudinal magnetoconductivity for a minimal model of TR broken untilted Weyl semimetal. (a) Increasing intervalley scattering strength results in strong sign-reversal. (b) In addition to this, infinitesimal strain now results in weak sign-reversal as well.  (c) When plotted as a function of the gauge field $B_5$, LMC is always strongly sign-reversed. (d) In the presence of an external magnetic field, we see signatures of weak-sign reversal as well. 
    In all the plots as we move from blue to the green curve we increase the intervalley scattering strength $\alpha$ from below $\alpha_c$ to above $\alpha_c$.}
    \label{fig:twonode_notilt_lmc_vs_B_vs_B5_vary_alpha}
\end{figure*}


\subsection{Longitudinal magnetoconductance \& strong and weak sign reversal}
Next, let us examine the behavior in the presence of an effective chiral gauge field ($B_5$) that may arise in inhomogeneous WSMs due to presence of strain. The chiral gauge field couples oppositely in opposite valleys, thus the net magnetic field becomes valley dependent, i.e., $B\rightarrow B+\chi B_5$. We first assume that $B_5$ is held parallel to the external magnetic field $B$. Fig.~\ref{fig:sigma001} plots the behavior of $\delta\sigma_{zz}(B)$, which is the change in LMC due to the magnetic field, i.e., $\delta\sigma_{zz}(B) = \sigma_{zz}(B) - \sigma_{zz}(B=0)$. We find that the increase or decrease of LMC depends on the direction of magnetic field, especially close to $B=0$. For example, when $\alpha<\alpha_c$, LMC decreases for positive values of magnetic field and increases for negative values of magnetic field. When $\alpha>\alpha_c$, the behavior is reversed. Furthermore, when $B$ is increased further away from zero  (in either direction), LMC increases (decreases) for both negative and positive values of B when $\alpha<\alpha_c$ ($\alpha>\alpha_c$).
Hence, it turns out that stating whether the longitudinal magnetoconductance is only positive or negative for a given scenario turns out to be rather ambiguous. 

To counter this, first we generalize the expression of magnetoconductivity to 
\begin{align}
\sigma_{zz}(B)= \sigma_{zz}^{(2)}(B-B_0)^2 + \sigma_{zz}^{(0)},
\label{eq:sigma_2}
\end{align}
The above definition allows us to shift the vertex of the parabola ($B_0$) away from origin, which is essential to fit the results presented in Fig.~\ref{fig:sigma001}. Now, in Fig.~\ref{fig:sigma001}(a) even though LMC is negative at low positive magnetic fields, it is in fact always positive when seen in reference to the vertex $B_0$, i.e., LMC is always positive when the change in the magnetic field and conductivity is seen with respect to the conductivity at $B_0$.
We call this as `weak' sign-reversal because the orientation of the parabola remains intact, and only the vertex is shifted from the origin, and also $\sigma_{zz}^{(2)}$ remains positive. Thus, when intervalley scattering is weak, strain in inhomogeneous WSMs drives the system to the `weak' sign-reversed state along a particular direction of the magnetic field. In summary, the characteristics defining weak sign-reversal are the following: (i) $B_0 \neq 0$, (ii) $\sigma_{zz}^{(0)} \neq \sigma_{zz}(B=0)$, (iii) $\mathrm{sign }\; \sigma_{zz}^{(2)}>0$. 

Now, when the strength of the intervalley scattering is greater than the critical value ($\alpha_c$), the orientation of the parabola is reversed, i.e., LMC does not again increase for $|B|>B_0$ unlike the earlier case, and $\sigma_{zz}^{(2)}$ becomes negative.
Due to this reason, we call this as `strong' sign-reversal. The only condition that we impose for strong sign-reversal is: (i)  $\mathrm{sign }\; \sigma_{zz}^{(2)}<0$, without any restriction to the values of $B_0$ and $\sigma_{zz}^{(0)}$. Therefore, the signatures of both strong and weak sign-reversal are:  (i) $B_0 \neq 0$, (ii) $\sigma_{zz}^{(0)} \neq \sigma_{zz}(B=0)$, (iii) $\mathrm{sign }\; \sigma_{zz}^{(2)}<0$. 
Since $B_0$ is shifted from the origin  due to infinitesimal strain even when $\alpha>\alpha_c$, we say that sufficiently strong intervalley scattering along with strain in inhomogeneous WSMs drives the system to show signatures of both weak and strong sign-reversal. This is demonstrated in Fig.~\ref{fig:sigma001} (b).
In general, the chiral gauge may be oriented away from the $z-$ axis and  rotated along the $xz$-plane. 
The variation of magnetoconductivity with respect to the angle $\gamma_5$ (the angle between $x$-axis and the $B_5$ field) is straightforward to understand. As $\gamma_5$ increases from zero to $\pi/2$, the contribution due to to the chiral gauge field increases in a sinusoidal fashion. We do not explicitly plot this behavior. 

In Fig.~\ref{fig:sigma002} we plot the parameters $B_0$ and $\sigma_{zz}^{(0)}$ as a function of the chiral gauge field and intervalley scattering strength. The transition from 'weak' to 'strong and weak' sign-reversed case (and vice-versa) is characterized by a sudden reversal in signs of the relative offset in conductivity $\sigma_{zz}^{(0)}$, as well as the vertex of the parabola $B_0$, i.e., $B_0\leq 0$ when $\sigma_{zz}^{(0)}\geq 0$, and vice-versa. In contrast, $\sigma_{zz}^{(2)}$ continuously interpolates across zero (not plotted). No discontinuity in $B_0$ or $\sigma^{(0)}$ is observed in the weak sign-reversed case, i .e., as the strain induced field is increased from zero for a constant intervalley scattering, the parameters $B_0$ and $\sigma_{zz}^{(0)}$ vary continuously. 

In Fig.~\ref{fig:twonode_notilt_lmc_vs_B_vs_B5_vary_alpha} we plot the the longitudinal magnetoconductivity as a function of magnetic field for different values of intervalley scattering. In the absence of chiral gauge field (Fig.~\ref{fig:twonode_notilt_lmc_vs_B_vs_B5_vary_alpha} (a)), as expected, we observe strong sign-reversal when $\alpha>\alpha_c$. In the presence of chiral gauge field field (Fig.~\ref{fig:twonode_notilt_lmc_vs_B_vs_B5_vary_alpha}(b)), we observe both strong and weak-sign reversal as also pointed out earlier. 
Typically, an increase in intervalley scattering strength decreases the magnetoconductivity, i.e., $|\sigma_{xz}(B,\alpha)|>|\sigma_{xz}(B,\alpha+\epsilon)|$, where $\epsilon$ is the infinitesimal increase in the scattering strength. We find this to be true even in the presence of strain induced chiral gauge field.
We particularly highlight this point here as this will be contrasted to the planar Hall conductivity that shows an anomalous increase in conductivity with increasing intervalley scattering strength. In Fig.~\ref{fig:twonode_notilt_lmc_vs_B_vs_B5_vary_alpha} (c) we plot the LMC in the presence of only chiral gauge magnetic field (i.e. $B=0$). Since, in this case the external magnetic field is zero, positive/negative LMC and weak/strong sign-reversal can only be defined with reference to the $B_5$ field. We find that the strain induced chiral gauge field by itself only results in strong sign-reversed phase irrespective of the intervalley scattering strength. We find this to be true even in the presence of external $B$-field (Fig.~\ref{fig:twonode_notilt_lmc_vs_B_vs_B5_vary_alpha} (d)).

\begin{figure}
    \centering
    \includegraphics[width=\columnwidth]{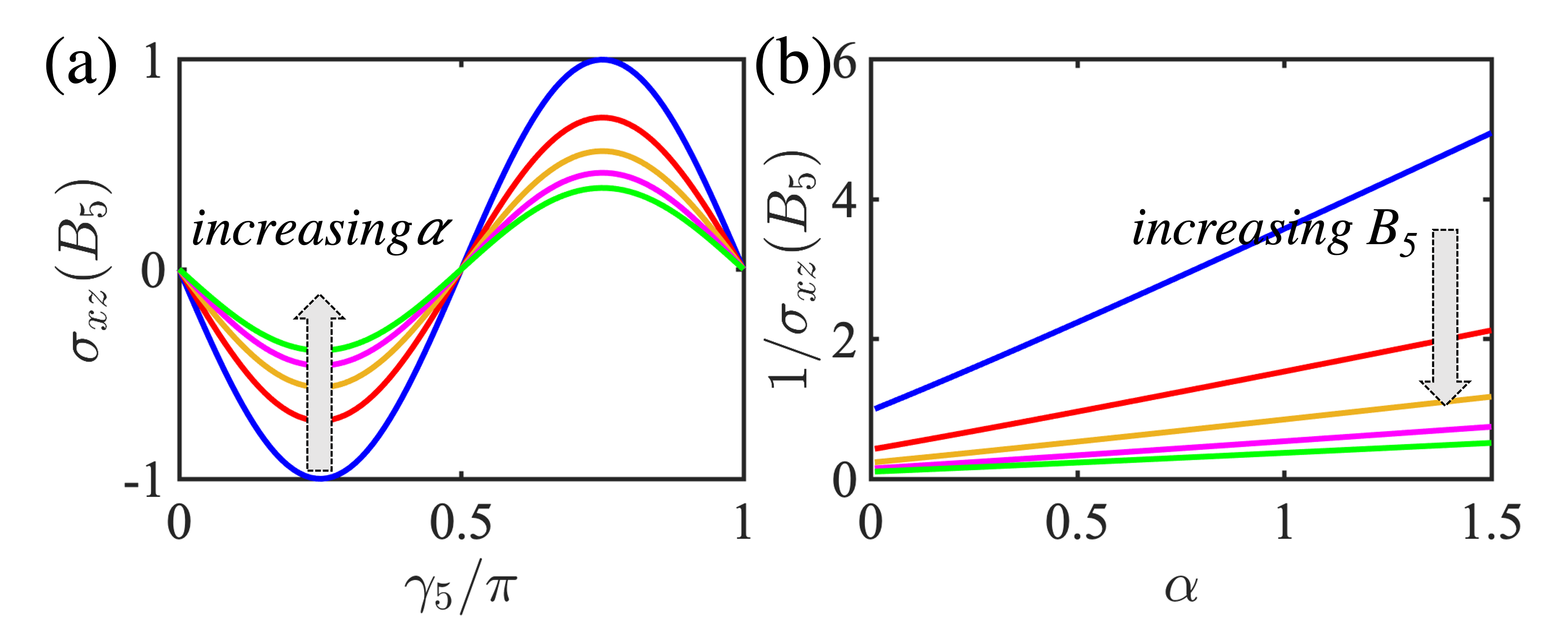}
    \caption{Planar Hall conductivity for a minimal model of untilted TR broken WSM in the absence of any magnetic field. (a) Variation with respect to the angle $\gamma_5$. Increasing $\alpha$ reduces the conductivity, as expected. (b) PHC behaves as the inverse of scattering strength. Since $\sigma_{xz}(B_5=0)=0$, we have normalized $\sigma_{xz}$ appropriately in both the plots. In creasing $B_5$ field increases the conductivity.}
    \label{fig:phc001}
\end{figure}
\begin{figure}
    \centering
    \includegraphics[width=\columnwidth]{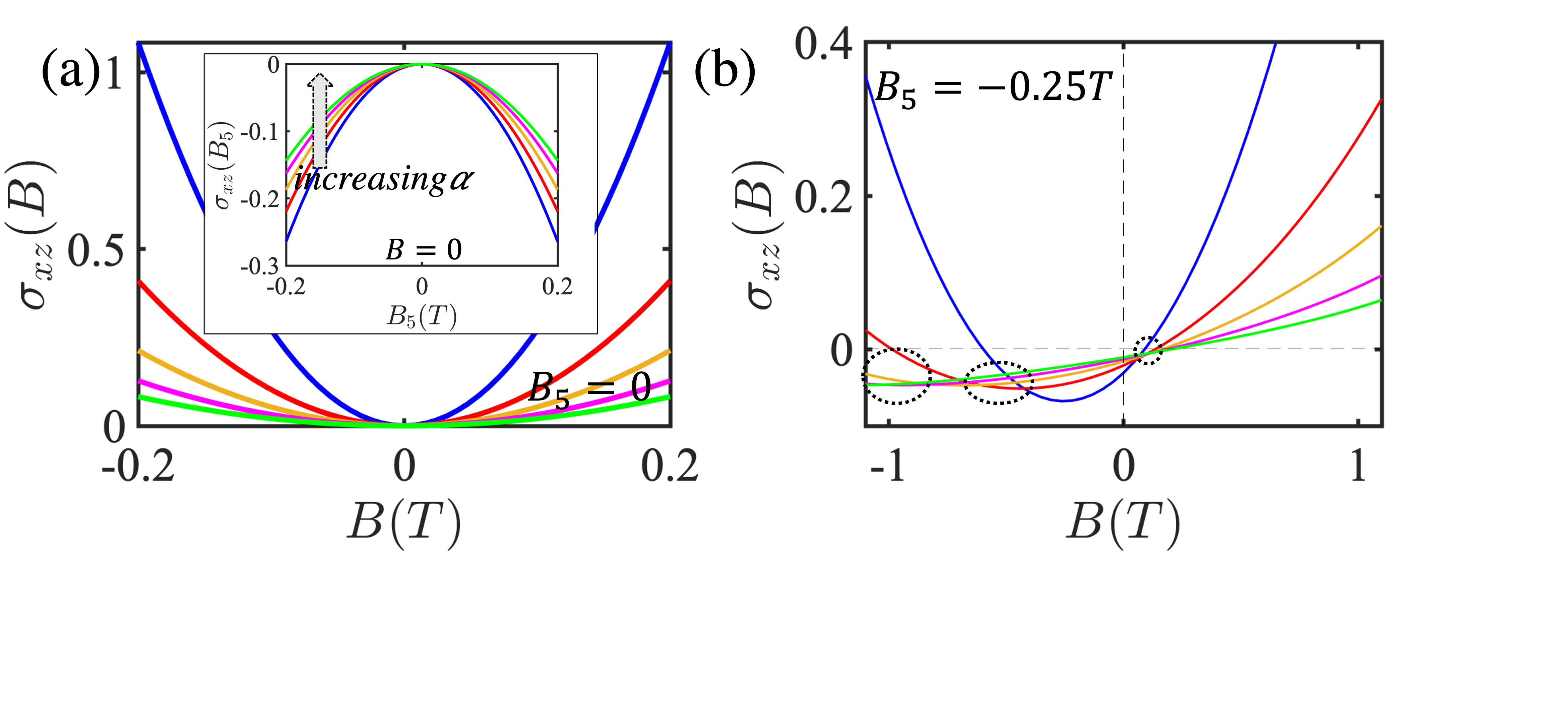}
    \caption{Planar Hall conductivity for a minimal model of untilted TR broken WSM. (a) PHC as a function of external magnetic field $B$ and no strain induced field ($B_5=0$) is compared with the inset where PHC has been plotted as a function of $B_5$ with no external field ($B=0$). The angle $\gamma$ was chosen to be equal to $\gamma_5$. Strain opposes the planar Hall effect albeit with different magnitude. (b) PHC in the presence of both magnetic field and strain. The chiral gauge field causes  weak sign-reversal. The dotted ellipses highlight regions that show an anomalous behavior with respect to intervalley scattering strength. The width of plots is reduced for better visibility. In all the curves, as we go from blue to green, we increase $\alpha$. All the plots are appropriately normalized.}
    \label{fig:phc002}
\end{figure}
\begin{figure}
    \centering
    \includegraphics[width=\columnwidth]{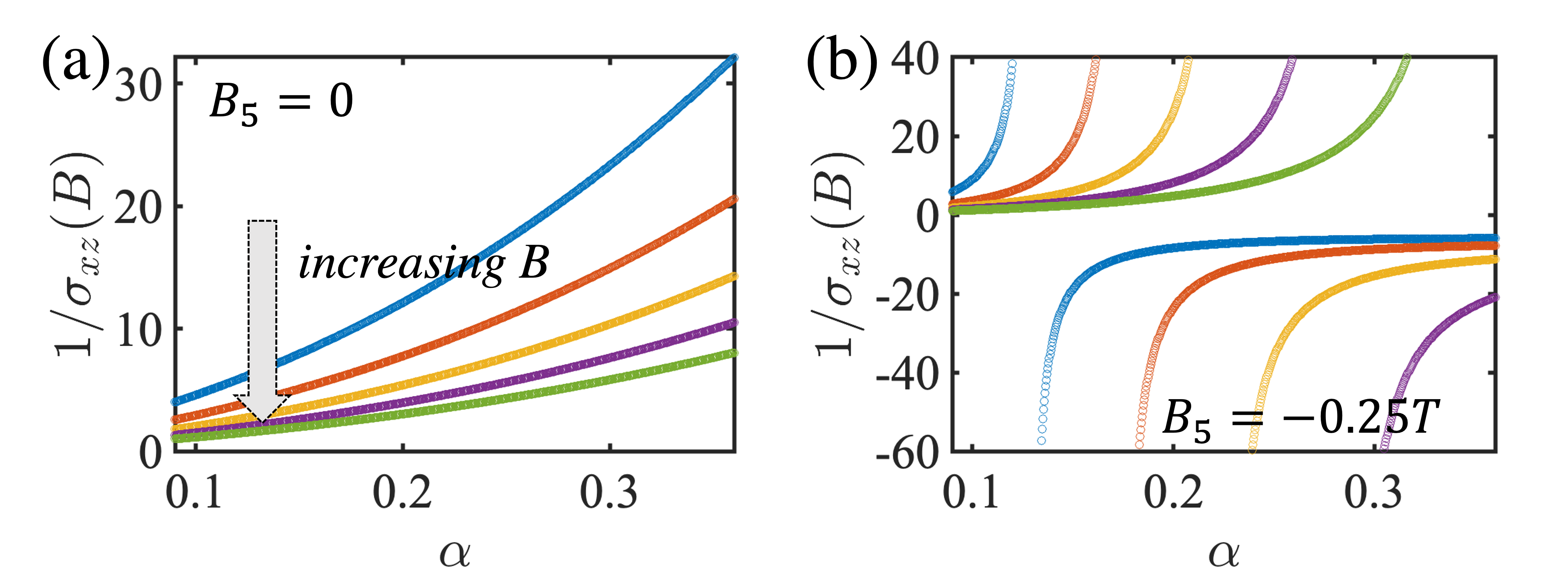}
    \caption{Planar Hall conductance for a minimal model of untilted WSM as a function of intervalley scattering strength. (a) in absence of $B_5$ field. (b) in presence of $B_5$ field. In all the curves, as we go from blue to green, we increase $B$. All the plots are appropriately normalized.}
    \label{fig:twonode_notilt_inv_sxz}
\end{figure}

\subsection{Planar Hall conductance}
Next, we study the effect of the chiral gauge field $B_5$ on the planar Hall conductance. The dependence on the magnetic field is typically quadratic and we may expand the planar Hall conductivity $\sigma_{xz}$ as
\begin{align}
\sigma_{xz}(B)= \sigma_{xz}^{(2)}(B-B_0)^2 + \sigma_{xz}^{(0)},
\label{eq:phc1}
\end{align} 
where $B_0$ is vertex of the parabola, and $\sigma_{xz}^{(2)}$ is the quadratic coefficient. 
The planar Hall conductivity depends on the angle of the applied magnetic field ($\sim \sin 2\gamma$), where $\gamma$ is the angle of the magnetic field with respect to the $x$-axis~\cite{nandy2017chiral}. To study the effect of strain, we first evaluate the planar Hall conductivity in the absence of any external magnetic field. In Fig.~\ref{fig:phc001} we plot the planar Hall conductivity $\sigma_{xz}(B_5)$ that is evaluated in the absence of external magnetic field. The angular behavior with respect to $\gamma_5$ is $\sim\sin 2\gamma_5$ as the case with the usual planar Hall conductivity. Here, we also explicitly examine the effect of intervalley scattering $\alpha$. Even though the conductivity is expected to decrease with increasing scattering, the functional form has still never been explicitly evaluated, especially when the scattering is momentum dependent. We numerically find that the planar Hall conductivity induced by the chiral gauge field  behaves as $\sim 1/\alpha$. 
\begin{figure}
    \centering
    \includegraphics[width=\columnwidth]{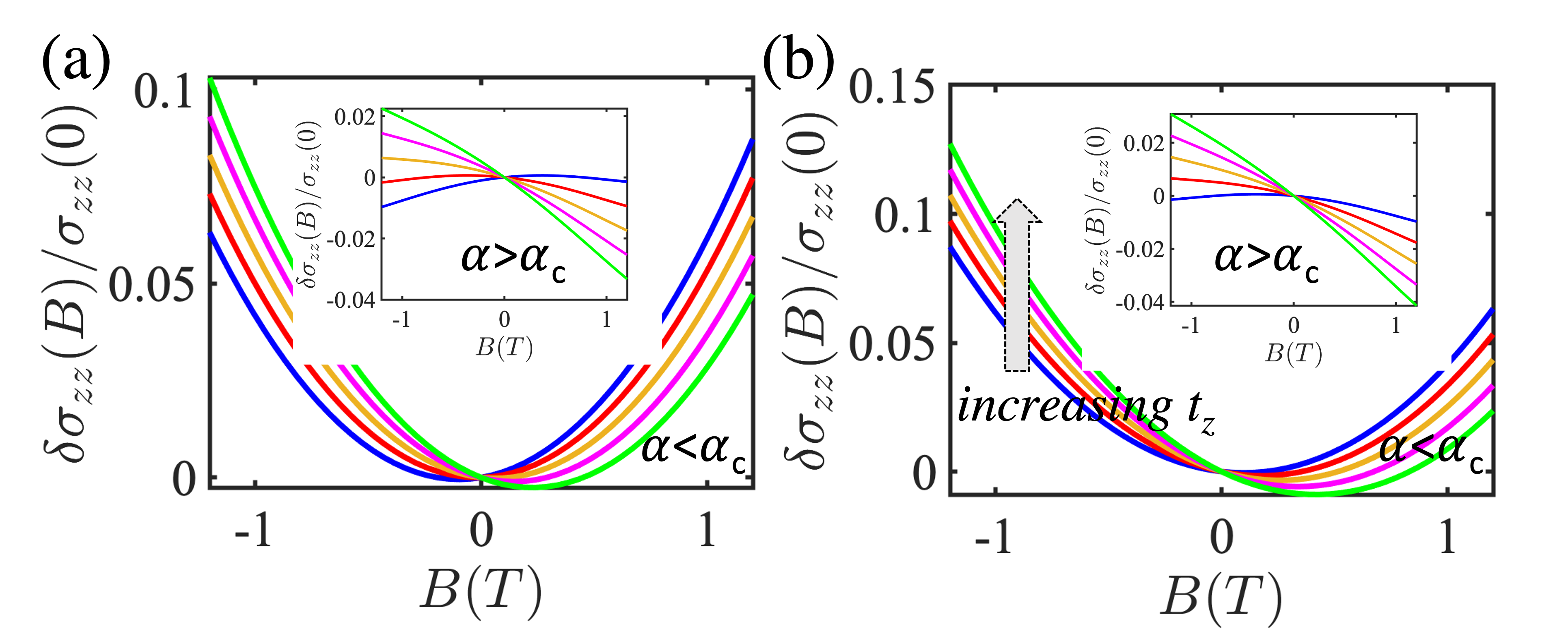}
    \caption{LMC for a tilted TR broken WSM (Eq.~\ref{Eq:HWeyl2}) with $t_z^{1} = -t_z^{(-1)}$. (a) When $B_5=0.1T$. (b) When $B_5=-0.1T$. The inset in both figures is for the case when $\alpha=1.2>\alpha_c$, while in the main figures $\alpha=0.2<\alpha_c$. As we move from blue to the green curve in both the plots, we increase $t_z/v_F$ from 0 to 0.06. The opposing effects and adding effects of strain and tilt are highlighted in (a) and (b) respectively.}
    \label{fig:tiltzLMC1}
\end{figure}
\begin{figure}
    \centering
    \includegraphics[width=\columnwidth]{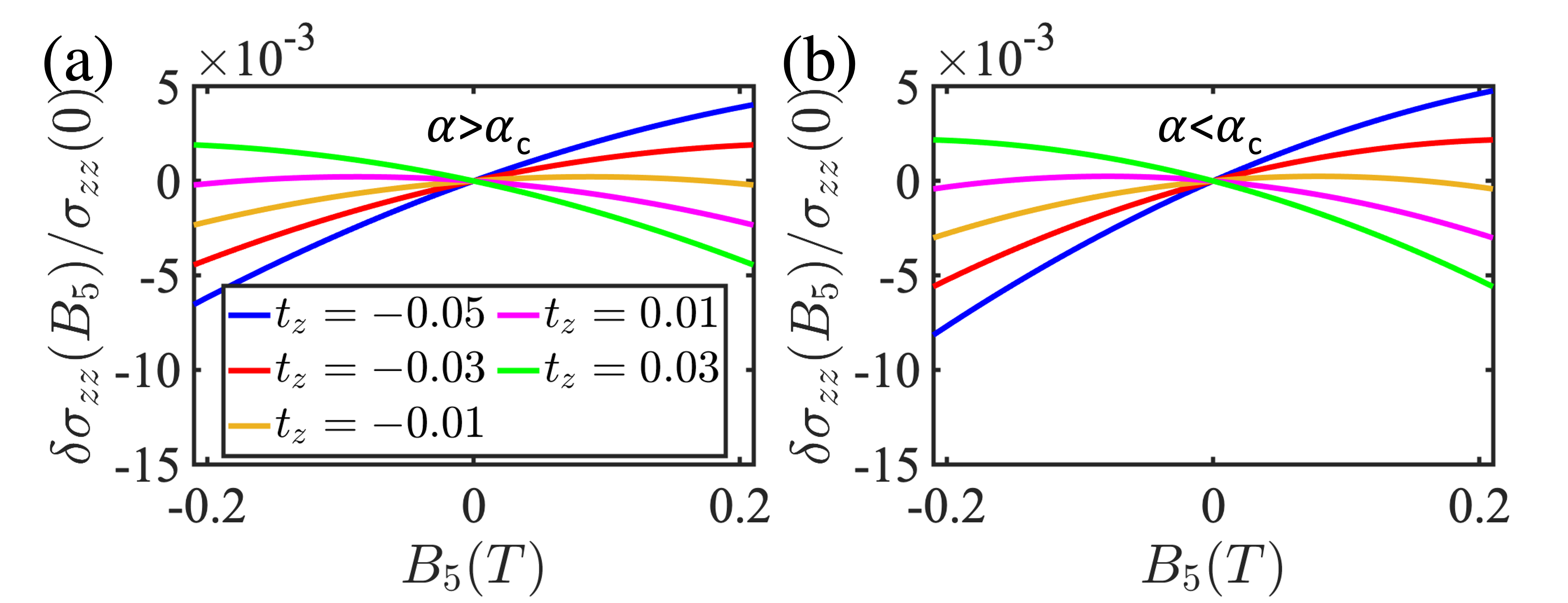}
    \caption{LMC for a tilted TR broken WSM when the tilts are oriented in the same direction. Both weak and strong sign-reversal is observed irrespective of the intervalley scattering strength. The legends are same in both the plots.}
    \label{fig:tiltzsameLMC2}
\end{figure}

We also compare and contrast the behavior of the planar Hall conductivity when (i) external magnetic field is applied and the strain induced field is absent, and (ii) when strain induced field is present but external magnetic field is absent. We find the contribution to the planar Hall conductivity to be different both in sign and magnitude, which is in contrast to earlier claims~\cite{ghosh2020chirality}. Specifically $\sigma_{xz} (B)$ increases with increasing $B$, while $\sigma_{xz} (B_5)$ decreases with increasing $B_5$.
This feature has been highlighted in Fig.~\ref{fig:phc002} (a). In other words, the chiral gauge field, alone, results in strong sign-reversal. We attribute this behavior to the inclusion of intervalley scattering, momentum dependence, as well as charge conservation that have been neglected in earlier works.

Finally, we also study the conductivity in the presence of both the external magnetic field and strain induced chiral magnetic field. In the presence of external magnetic field, the effect of strain is to shift and tilt the conductivity parabola, thereby resulting in weak sign-reversal of the conductivity as shown in Fig.~\ref{fig:phc002} (b). In contrast to the longitudinal magnetoconductivity, PHC never shows strong sign-reversal even on increasing the intervalley scattering above the critical value. However, interestingly, we find that in a certain window of the magnetic field, increasing intervalley scattering strength increases the magnitude of the planar Hall conductivity, which is counter-intuitive. We understand this behavior due to the opposing effects of strain induced PHC and magnetic field induced PHC. As discussed before, both of them individually have opposite and unequal contributions to the planar Hall conductivity. This is better visualized in Fig.~\ref{fig:twonode_notilt_inv_sxz}, where we plot the planar Hall conductivity as a function of the intervalley scattering strength $\alpha$. First, we notice that in the absence of $B_5$-field, the Hall conductivity shows some amount of non-linearity as a function of $1/\alpha$. This is contrasted to Fig.~\ref{fig:phc001}(b) (the case when $B=0$, $B_5\neq 0$) where linear behavior was observed for all ranges of $\alpha$. Second, in the presence of $B_5$ field, the behavior of $\sigma_{xz}$ with respect to $\alpha$ can be strikingly different. Due to the weak sign reversal, $\sigma_{xz}$ can switch sign, which explains the divergences in the plot in Fig.~\ref{fig:twonode_notilt_inv_sxz} (b). Furthermore, we find that when $\sigma_{xz}$ switches sign from positive to negative, the behavior with respect to $\alpha$ becomes anomalous, i.e., increasing $\alpha$, increases the magnitude of $\sigma_{xz}$. Such an anomalous behavior with respect to the intervalley scattering strength is not observed for longitudinal magnetoconductivity.

\subsection{Time-reversal broken WSM with tilt}
Having discussed the physics of strain induced gauge field in a minimal untilted model of Weyl fermions, we now discuss the case when there is a finite tilt in the Weyl cones. 
The Hamiltonian is given by 
\begin{align}
    H = \sum\limits_\chi \sum\limits_\mathbf{k} {\chi\hbar v_F \left( \mathbf{k}\cdot\boldsymbol{\sigma} + t^\chi_z k_z\right)}
    \label{Eq:HWeyl2}
\end{align}

\begin{figure}
    \centering
    \includegraphics[width=\columnwidth]{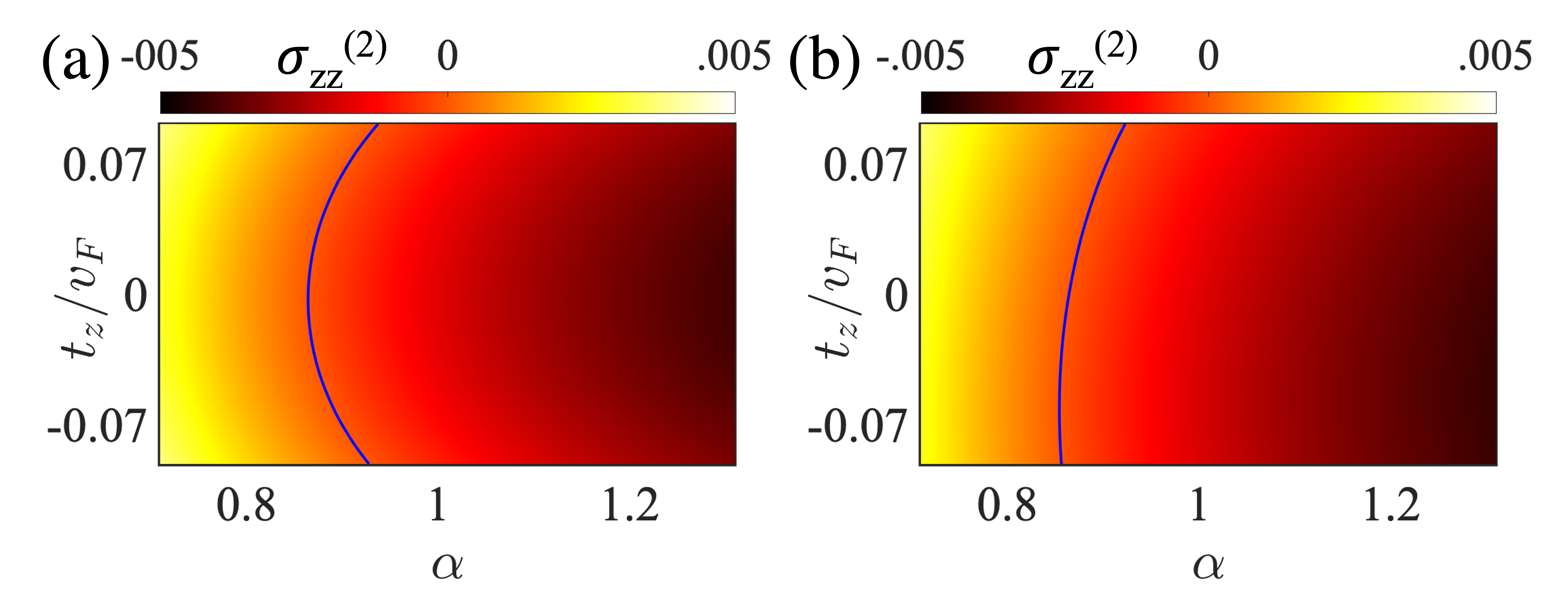}
    \caption{(a) The quadratic coefficient of the longitudinal magnetoconductivity $\sigma_{zz}^{(2)}$ for tilted TR broken WSM. (a) $t_z^{(1)} = -t_z^{(-1)}$. (b) $t_z^{(1)} = t_z^{(-1)}$. Strain induced chiral magnetic field was fixed to $B_5=0.1T$ in both the cases. The blue contour separates the regions when $\sigma_{zz}^{(2)}>0$ and when $\sigma_{zz}^{(2)}<0$ (strong sign-reversal).}
    \label{fig:tiltz_same_opp_lmccolorplot1}
\end{figure}

\begin{figure*}
    \centering
    \includegraphics[width=2\columnwidth]{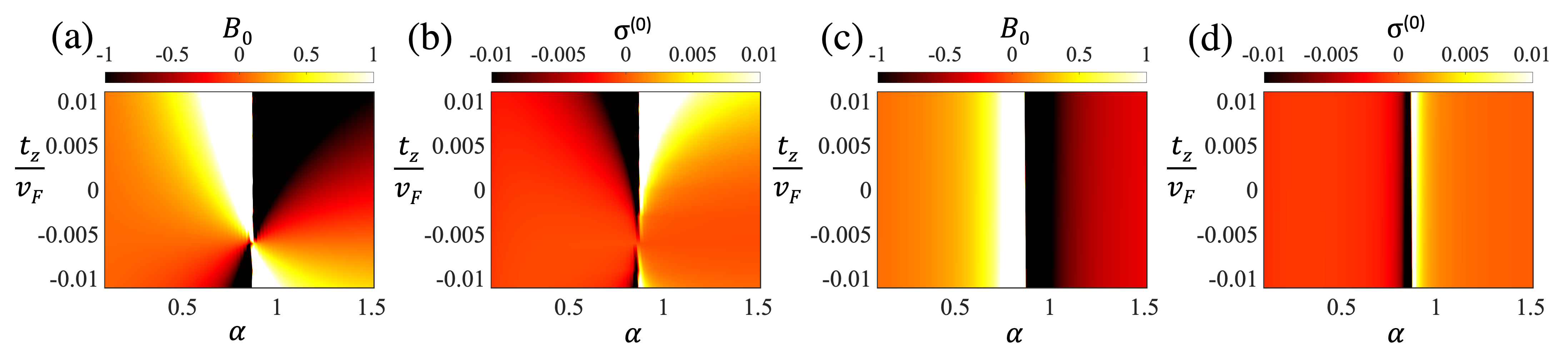}
    \caption{LMC parameters for tilted TR broken WSMs. The center of the parabola $B_0$ (a) and $\sigma^{(0)}$ as a function of the tilt parameter and intervalley scattering strength, in the presence of a fixed value of chiral gauge magnetic field $B_5=0.1T$. The tilts are oriented opposite to each other in plots (a) and (b). The plots (c) and (d) are for the case when the Weyl cone tilts are oriented in the same direction.}
    \label{fig:twonode_tiltz_colorplot_B_sigatB0}
\end{figure*}

Here $t_z$ is the tilting parameter along the $z$-axis. We only focus on the case when $t^\chi_{z}<v_F$, thus restricting ourselves to type-I Weyl semimetals. 
Depending on whether the two cones are tilted along the same or opposite direction, the behavior of both LMC and PHC can be different. In the absence of strain, if the cones are tilted in opposite directions, i.e., $t^\chi_z=-t^{\chi'}_z$, a linear in magnetic field term is added to the overall longitudinal magnetoconductivity, and the parabola is shifted and tilted along a particular direction. In other words, we can say that tilting results in weak sign-reversal, although this has never been explicitly pointed out in  earlier works~\cite{sharma2017chiral, das2019linear, ahmad2021longitudinal}. When the intervalley scattering strength is large, tilting the Weyl cones results in both weak and strong sign-reversal.
In the presence of both tilt and strain, we arrive at a very interesting scenario. Both of these parameters, i.e., $t_z$ and $B_5$, can tilt the LMC parabola either in the same direction or opposite direction, and this depends on the angle between the tilt direction and the strain induced gauge field. In Fig.~\ref{fig:tiltzLMC1} we plot the longitudinal magnetoconductivity for a tilted TR broken WSM presented in Eq.~\ref{Eq:HWeyl2} when the Weyl cones are tilted opposite to each other. Depending on the direction of the strain induced gauge field $B_5$, the effects of tilting and strain can either add up or even cancel out. In Fig.~\ref{fig:tiltzLMC1} (a), $B_5>0$, and the strain and tilting effects work in opposite directions, while in Fig.~\ref{fig:tiltzLMC1} (b), $B_5<0$, and the strain and tilting effects work in the same direction.

\begin{figure}
    \centering
    \includegraphics[width=\columnwidth]{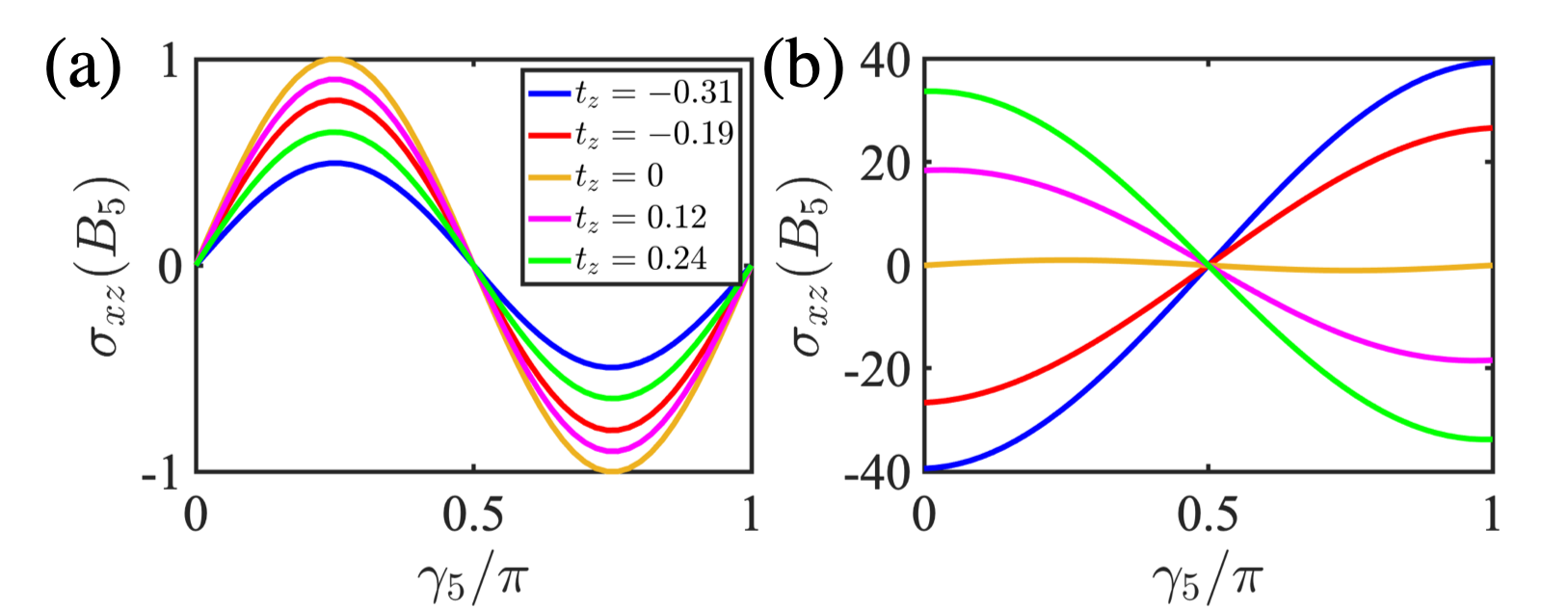}
    \caption{(a) The planar Hall conductance in TR broken tilted WSM as a function of the angle $\gamma_5$ when (a) the cones are tilted along opposite direction, and (b) cones are oriented along the same direction. The legends are the same in both the plots. Both plots are appropriately normalized such that the yellow curve is identical in both the figures as expected.}
    \label{fig:twonodes_tiltz_opp_same_phc_vs_gm5}
\end{figure}

\begin{figure}
    \centering
    \includegraphics[width=\columnwidth]{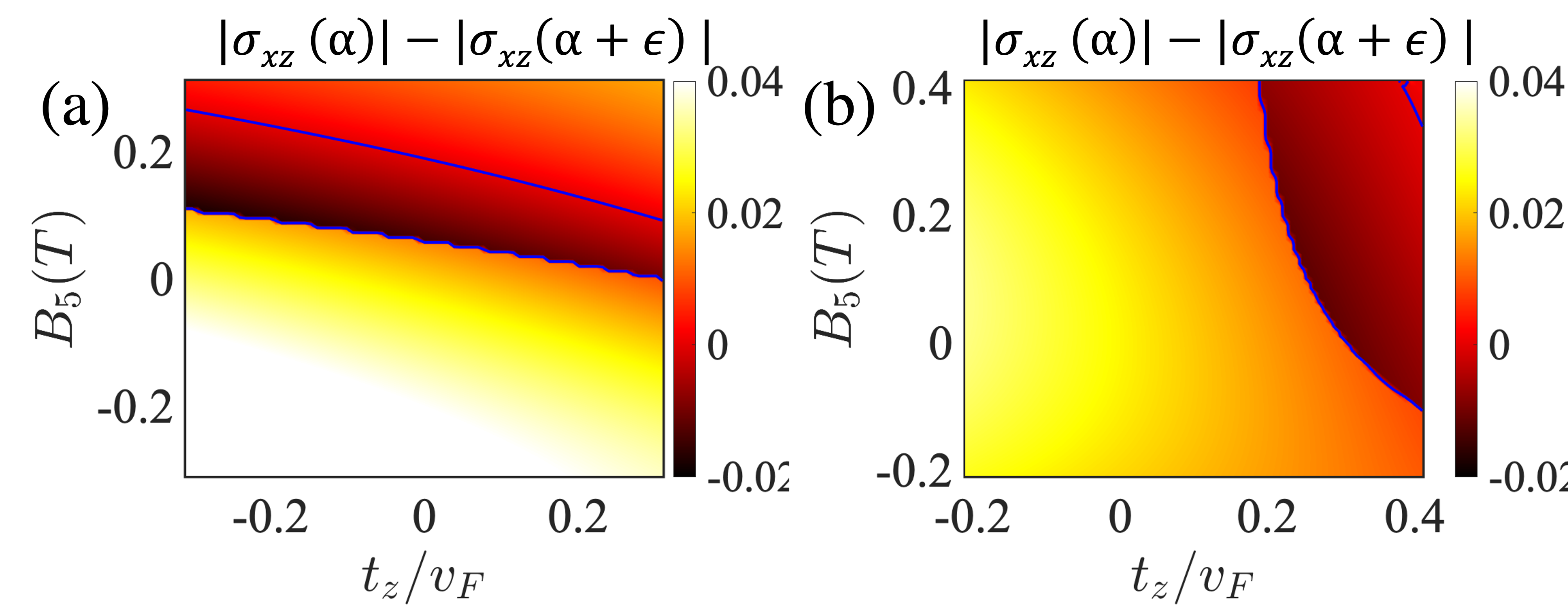}
    \caption{Change in the magnitude of the planar Hall conductivity ($|\sigma_{xz}(\alpha)|-|\sigma_{xz}(\alpha+\epsilon)|$) for a tilted TR broken WSM (Eq.~\ref{Eq:HWeyl2}) on infinitesimally increasing in the scattering strength (by $\epsilon$). (a) the Weyl cones are tilted in opposite direction. (b) the Weyl cones are tilted in the same direction. In the region enclosed within blue contours, we find anomalous behavior of conductivity with the scattering strength, i.e., the magnitude of the conductivity increases on the increase of scattering strength. We choose $\alpha=0.5$, and $\epsilon=0.01$.}
    \label{fig:twonodes_tiltz_opp_same_color_deltaphc}
\end{figure}

\begin{figure*}
    \centering
    \includegraphics[width=2\columnwidth]{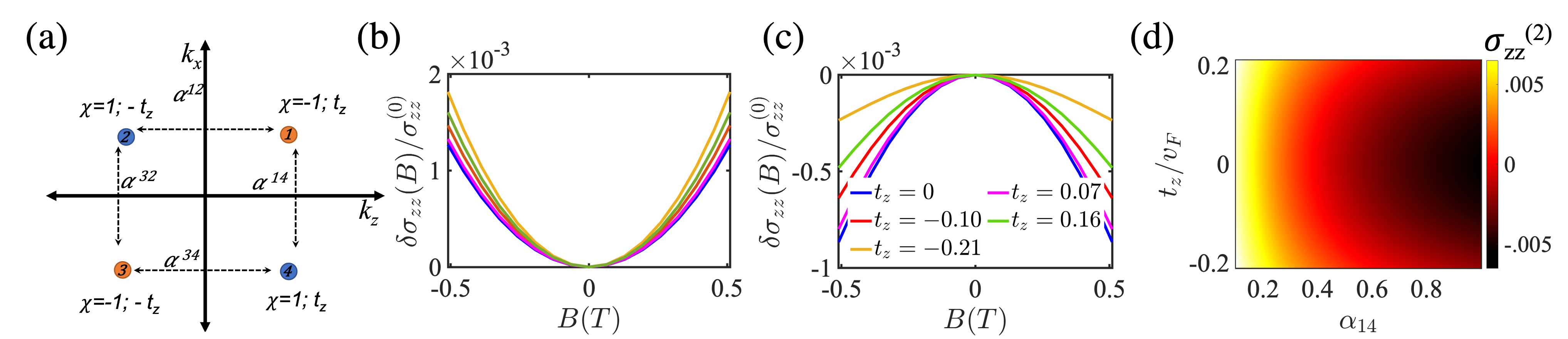}
    \caption{(a) Schematic of Weyl nodes in a prototype model of an inversion asymmetric Weyl semimetal. Here $\chi$ is the chirality, $t_z$ is the tilt, and $\alpha^{ij}$ are scattering rates from node $i$ to node $j$. (b) LMC as a function of magnetic field when the intervalley scattering rates are less than the critical value. (c) LMC as a function of magnetic field when the intervalley scattering rates are above the critical value. The legends in (b) and (c) are identical. (d) $\sigma_{zz}^{(2)}$ for a fixed value of $\alpha_{12}=0.19$. Plots (b), (c), and (d) are in the absence of strain, i.e., $B_5=0$.}
    \label{fig:fournodelmc1}
\end{figure*}
\begin{figure*}
    \centering
    \includegraphics[width=2\columnwidth]{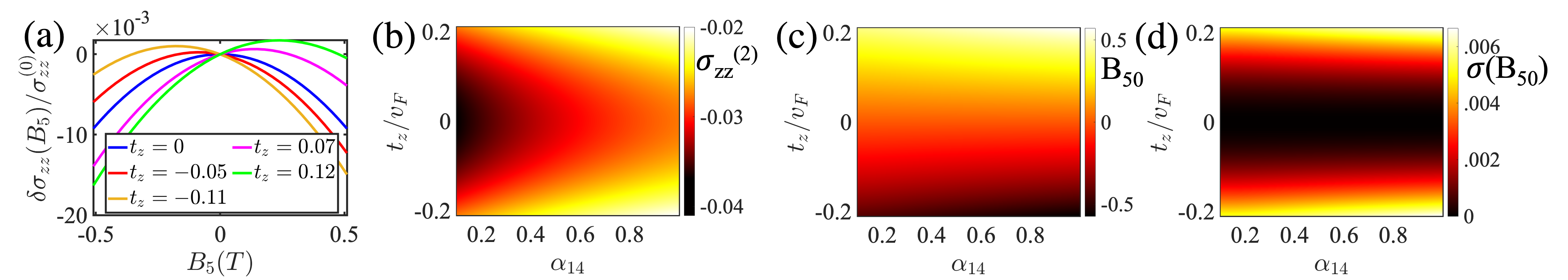}
    \caption{LMC for inversion asymmetric Weyl semimetal in the presence of strain induced chiral magnetic field ($B_5$) but absence of magnetic field. (a) A finite tilt can result in weak sign-reversal. The plot is for a fixed value of $\alpha_{12}=0.4$, but the qualitative behavior is independent of scattering strength. (b), (c), and (d) plot the parameters $\sigma_{zz}^{(2)}$, $B_{50}$, and $\sigma(B_{50})$ as a function of parameters $\alpha_{14}$ and $t_z$. We fixed $\alpha_{12}=0.19$.}
    \label{fig:fournodelmc2}
\end{figure*}
\begin{figure}
    \centering
    \includegraphics[width=\columnwidth]{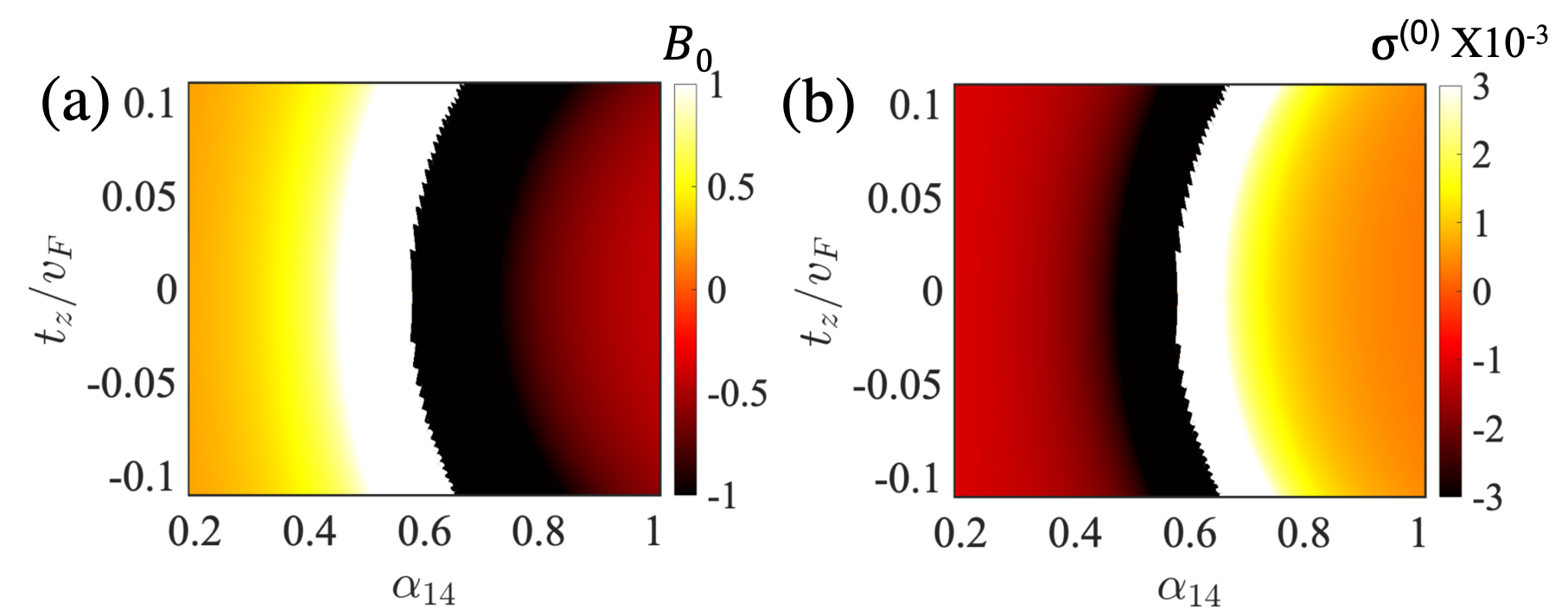}
    \caption{The parameters $B_0$ (a) and $\sigma^{(0)}$ (b) for inversion asymmetric Weyl semimetals (Eq.~\ref{eq_H4nodes}). We have fixed $\alpha_{12}=0.3$, $B_5=0.1T$. Weak sign reversal is not observed and strong sign-reversal occurs at $\alpha_{14}=\alpha_{14c}(t_z)$. }
    \label{fig:fournodes_lmc_colorplot_B0_sigatB0_1}
\end{figure}
\begin{figure*}
    \centering
    \includegraphics[width=2\columnwidth]{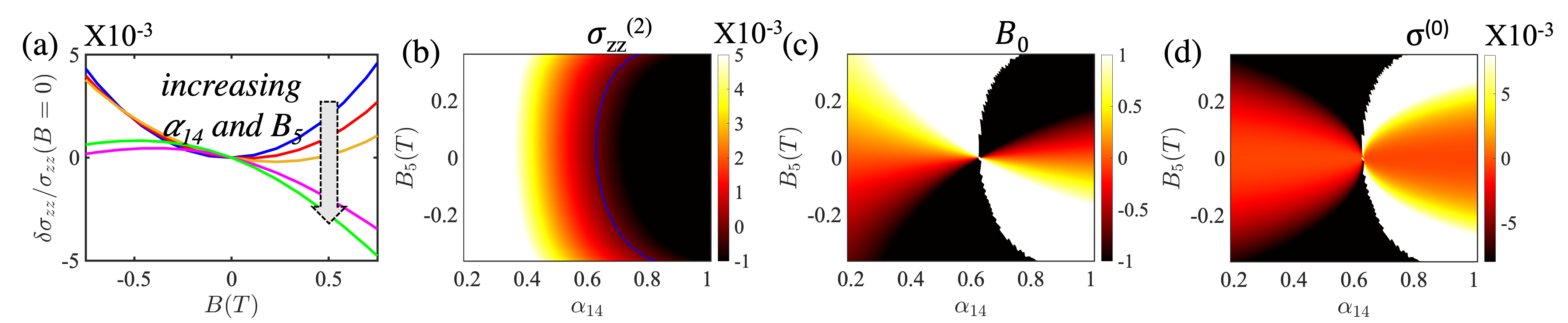}
    \caption{(a) LMC for inversion asymmetric Weyl semimetal. As we move from the blue to the green curve, we simultaneously increase $B_5$ as well as $\alpha_{14}$. Both weak and strong sign-reversal are exhibited. The plots (b), (c), and (d) plot the parameters $\delta\sigma_{zz}^{(2)}$, $B_0$, and $\sigma^{(0)}$ for fixed $\alpha_{12}$ and $t_z\neq 0$. The blue contour in plot (b) separates the phases where $\sigma_{zz}^{(2)}$ changes sign. Again, we see signatures of both weak and strong sign-reversal. The tilt parameter is fixed to $t_z/v_F=-0.1$.}
    \label{fig:fournodes_lmc_colorplot_varyB5_vary_a14}
\end{figure*}

\begin{figure}
    \centering
    \includegraphics[width=\columnwidth]{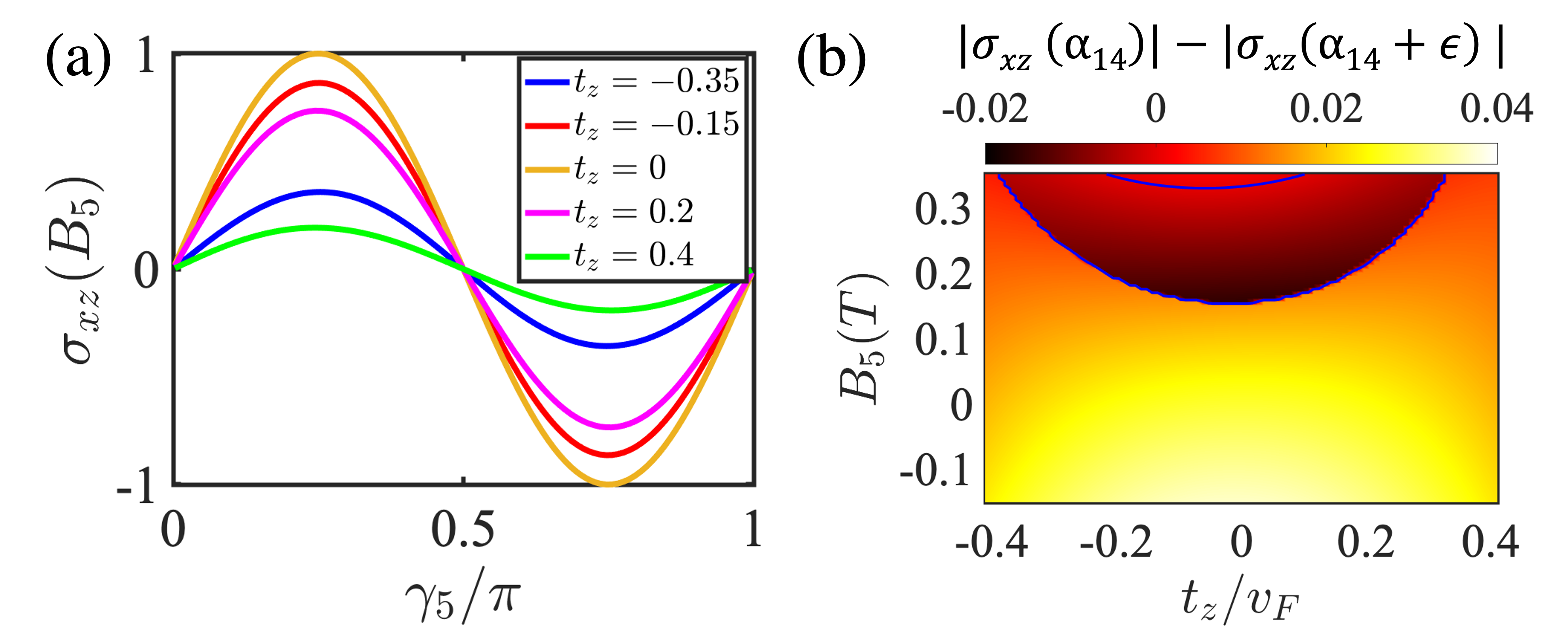}
    \caption{Planar Hall conductance for inversion asymmetric Weyl semimetal. (a) PHC as a function of $\gamma_5$, when $B=0$, and $B_5\neq 0$. (b) The change in the magnitude of the planar Hall conductivity on increasing $\alpha_{14}$ infinitesimally. In the region between the blue contours, we observe anomalous increase in conductivity. Here we fix, $B=1T$, $\alpha_{12}=0.4$, $\alpha_{14}=0.5$, and $\epsilon=0.01$. }
    \label{fig:fournodes_sxz_colorplot}
\end{figure}

In the absence of strain it is known that when the nodes are oriented along the same direction ($t^1_z = t^{-1}_z$), the linear component of the longitudinal magnetoconductivity does not survive as the contributions from both nodes cancel out~\cite{sharma2017chiral, das2019linear, ahmad2021longitudinal}. Hence, as expected, only strong sign-reversal is observed as a function of intervalley scattering strength. Now, in the presence of only strain induced field, such cancellation does not occur and one observes weak sign-reversal as a function of the tilt parameter. Furthermore, in the presence of $B_5$-field and absence of external magnetic field, we observe both strong and weak sign-reversal. To illustrate this, in Fig.~\ref{fig:tiltzsameLMC2} we plot LMC for a tilted TR broken WSM when the tilts are oriented in the same direction. When both magnetic field and strain induced chiral magnetic field are present, the combination of two can give rise to interesting features. In Fig.~\ref{fig:tiltz_same_opp_lmccolorplot1} we plot the quadratic coefficient $\sigma_{zz}^{(2)}$ as a function of both tilt and intervalley scattering strength in the presence of a $B_5$ field. We note that the presence of the tilt parameter curves the contour $\alpha_c$ separating the two strong sign-reversed regions, i.e., $\alpha_c = \alpha_c(t_z)$. The curvature is different when the Weyl cones are oriented opposite to each other or oriented along the same direction. 

Similarly, very striking features are observed for the parameters $B_0$ (the vertex of the parabola) as well as $\sigma_{zz}^{(0)}$. We demonstrate this in Fig.~\ref{fig:twonode_tiltz_colorplot_B_sigatB0}. We fix strain induced gauge field to be around $B_5=0.1T$. Let us first focus on the case when the Weyl cones are oriented opposite to each other.  When $\alpha<\alpha_c$, the sign of $B_0$ changes continuously from negative to positive as $t_z$ is varied from negative to positive. On the other hand, when $\alpha>\alpha_c$, the sign of $B_0$ changes from positive to negative as $t_z$ is varied from negative to positive. The effects of strain and tilt and strain can either add up or cancel out and the combination can tilt the parabola overall to the left or to the right resulting in weak sign-reversal. 
This is  demonstrated in the color plot  in Fig.~\ref{fig:twonode_tiltz_colorplot_B_sigatB0} (a). When $\alpha>\alpha_c$, the sign of $B_0$ changes discontinuously (feature of strong sign reversal).
Now, since weak sign-reversal does not change the sign of $\sigma^{(0)}$, we do not see a sign change in $\sigma^{(0)}$ as one varies the tilt for a given value of $\alpha$. The sign change in $\sigma^{(0)}$ only occurs as a result of strong sign-reversal (Fig.~\ref{fig:twonode_tiltz_colorplot_B_sigatB0} (b)). Now, when the cones are oriented along the same direction, the linear component arising from the tilt is canceled out and hence we do not observe any change in $B_0$ or $\sigma^{(0)}$ by varying the tilt. The only change occurs at $\alpha=\alpha_c$ due to strong sign-reversal. This is highlighted in Figs.~\ref{fig:twonode_tiltz_colorplot_B_sigatB0} (c) and (d). 

Next we discuss the strain induced planar Hall effect for tilted TR broken Weyl semimetals. 
When the cones are oriented along the opposite directions we observe a $\sim\sin 2\gamma_5$ behavior and the effect of the tilt is only quantitative, and so is the effect of varying intervalley scattering strength. On the other hand, when the cones are oriented along the same direction, the behavior changes to $\sim\sin\gamma_5$. Changing the tilt parameter can switch the sign of the planar Hall conductance as well, and result in qualitative changes in the behavior while changing the intervalley scattering strength only changes the overall magnitude. We demonstrate these features in Fig.~\ref{fig:twonodes_tiltz_opp_same_phc_vs_gm5}. 

Finally, we discuss the behavior of conductivity on changing the intervalley scattering strength $\alpha$. In Fig.~\ref{fig:twonodes_tiltz_opp_same_color_deltaphc}, we plot the change in the magnitude of the planar Hall conductivity ($|\sigma_{xz}(\alpha)|-|\sigma_{xz}(\alpha+\epsilon)|$) for an  infinitesimal increase in the scattering strength (by a small amount $\epsilon$). In both cases, i.e., when the Weyl cones tilted in opposite direction, and when the Weyl cones are tilted in the same direction, we find regions in the $B_5-t_z$ space where anomalous behavior of the Hall conductivity is observed, i.e., the magnitude of conductivity increases on increasing the intervalley scattering strength. We had already seen this behavior for untilted WSM as well (Fig.~\ref{fig:twonode_notilt_inv_sxz}), and here we calculate its dependence on the tilting of the Weyl cones. 
Before closing this section, we point out that in experiments where strain can be applied and manipulated on the inhomogeneous samples can test the above predictions.

\section{Inversion asymmetric Weyl semimetals}
Having discussed the effect of strain in time-reversal broken WSMs we now move on to the case of inversion asymmetric WSMs. To this end, we will restrict our attention to the following minimal model for an inversion asymmetric WSM that consists of four nodes as dictated by symmetry considerations:
\begin{align}
    H = \sum\limits_{n=1}^4 \left(\chi_{n}\hbar v_F \mathbf{k}\cdot\boldsymbol{\sigma} + \hbar v_F t_z^n k_z \sigma_0\right).
    \label{eq_H4nodes}
\end{align}
The system consists of four Weyl nodes located at the points $\mathbf{K}=(\pm k_0,0,\pm k_0)$ in the Brillouin zone. In Eq.~\ref{eq_H4nodes}, $\chi_n$ is the chirality, and we are also introducing the parameter $t_z^n$, that represents the tilting of the Weyl cone. The Weyl cones are assumed to be tilted only along the $z$ direction. Specifically, $(1,t_z)$=$(\chi_1,t_z^{(1)})=(-\chi_2,t_z^{(2)})=(\chi_3,-t_z^{(3)})=(-\chi_4,-t_z^{(4)})$, such that inversion symmetry is broken. The tilt parameter $t_z$ is considered to be less than unity. Fig.~\ref{fig:fournodelmc1}(a) plots the schematic diagram of this prototype inversion asymmetric Weyl semimetal. 
Specifically, we must consider four intranode scattering channels (node $n\Longleftrightarrow n$) and four internode scattering channels (node $n\Longleftrightarrow [n+1] \text{mod } 4$). The dimensionless scattering strength between node $m$ and node $n$ is denoted as $\alpha^{mn}$. For simplicity, we ignore the scattering between nodes (4 $\Longleftrightarrow$ 2) and nodes (1 $\Longleftrightarrow$ 3) since they involve a large momentum transfer compared to others. The four internode scatterings can be divided into two categories: (i) scattering between Weyl cones of opposite chirality and opposite tilt orientation (1 $\Longleftrightarrow$ 2) and (3 $\Longleftrightarrow$ 4), and (ii) scattering between Weyl cones of opposite chirality and same tilt orientation (1 $\Longleftrightarrow$ 4) and (2 $\Longleftrightarrow$ 3). Since both these categories result in different behaviors, it is of interest to see the interplay between the two. 
We first examine the behavior of longitudinal magnetoconductivity in the absence of any strain. Earlier, we examined that for a system of only two tilted cones (of opposite chirality), `weak' sign-reversal is possible only if the cones are oriented opposite to each other. However, in the current case, `weak' sign-reversal generated by internode scattering channel (1 $\Longleftrightarrow$ 2) is exactly cancelled by scattering channel (4 $\Longleftrightarrow$ 3). Second, the scattering (1 $\Longleftrightarrow$ 4) and (2 $\Longleftrightarrow$ 3) do not cause weak sign reversal as they involve Weyl cones with the same tilt. Therefore, in the absence of $B_5$ field, weak sign-reversal is not observed for the case of an inversion asymmetric WSM. 
In Fig.~\ref{fig:fournodelmc1} we plot longitudinal magnetoconductivity  for the inversion asymmetric Weyl semimetal (Eq.~\ref{eq_H4nodes}) in the absence of strain induced chiral gauge field $B_5$. As discussed, we do not observe any signature of weak sign-reversal, and there is only strong sign-reversal when $\alpha_{12}$ and/or $\alpha_{14}$ are large enough. Increasing tilt does not qualitatively change the behavior and increasing the magnitude of the  tilt in either direction is  only seen to increase the magnitude of magnetoconductivity. 

Next, we study the behavior in the absence of external magnetic field but in presence of strain induced gauge field $B_5$. First, similar to the case with TR broken Weyl semimetals, we find that strain induced chiral magnetic field $B_5$ always results in a negative LMC coefficient $\sigma_{zz}^{(2)}$. This results in in contradiction to earlier claims that find an increase in longitudinal magnetoconductivity with strain~\cite{grushin2012consequences,ghosh2020chirality}. The reason can be traced out to the non-inclusion of intervalley scattering, momentum dependent scattering, and charge conservation, all of which are included in the current work (see Appendix A).
Furthermore, we find that strain, by itself results in strong sign-reversal, while tilting results in weak sign reversal.
In Fig.~\ref{fig:fournodelmc2} (a) we plot LMC as a function of strain induced magnetic field $B_5$, which clearly demonstrates these features.
As before, we fit the magnetoconductivity via the following expression 
\begin{align}
\sigma_{zz}(B_5)= \sigma_{zz}^{(2)}(B-B_{50})^2 + \sigma_{zz}(B_{50}),
\label{eq:sigma_25}
\end{align}
where the slope of the conductivity $\sigma_{zz}^{(2)}$ is always found to be negative irrespective of the value of tilt, strain, intervalley scattering strengths across either nodes. The center of the parabola ($B_{50}$) directly correlates with the tilt parameter $t_z$. Depending on the sign of $t_z$, $B_{50}$ can be either positive or negative. The parameter $B_{50}$ is also found to have dependence on the scattering strength, but this dependence is relatively weak compared to the dependence on $t_z$.  In Figs.~\ref{fig:fournodelmc2} (b), (c), and (d), we plot the parameters $\sigma_{zz}^{(2)}$, $B_{50}$, and $\sigma_{zz}(B_{50})$ as a function of $\alpha_{14}$, and $t_z$, keeping $\alpha_{12}$ fixed, and $B=0$. No sharp discontinuities are observed in the parameters since the system is already in strong sign-reversed state.

In inversion asymmetric inhomogeneous Weyl semimetals, interesting effects can occur as a result of the interplay between the strain induced chiral gauge field, external magnetic field, and the tilt parameter. To study the same, we examine LMC as a function of external magnetic field for a fixed value of chiral gauge field, and use Eq.~\ref{eq:sigma_2} to evaluate the fit parameters $B_0$, $\sigma_{zz}^{(2)}$, and $\sigma_{zz}^{(0)}$. We do not find a signature weak sign-reversal, and only strong sign-reversal occurs when the intervalley scattering $\alpha_{14}>\alpha_{14c}$, where $\alpha_{14c}$ now is a function of tilt parameter.  Around $\alpha=\alpha_{14c}(t_z)$ we find a sharp change in the sign of the parameters $B_0$ and $\sigma_{zz}^{(0)}$ that corresponds to a continuous change of sign in $\sigma_{zz}^{(2)}$ as well. 
It is worthwhile pointing that by identifying the parameters $B_0$ and $\sigma_{zz}^{(0)}$ from the experimentally measured conductivity, their signs may help identify the dominant scattering mechanisms in the system, i.e., either internode or intranode scattering, and also provide us insight about the strain in the samples as well as the tilting if the Weyl cones.

Experimentally, one may also study LMC in inversion asymmetric Weyl semimetals by tuning the amount of strain in the system. Therefore it is of interest to study the effect of varying strain on LMC. 
In Fig.~\ref{fig:fournodes_lmc_colorplot_varyB5_vary_a14} (a) we plot $\delta\sigma_{zz} = \sigma_{zz}(B) - \sigma_{zz}(B=0)$ simultaneously varying the intervalley scattering strength $\alpha_{14}$ as well as the strain induced chiral gauge field $B_5$. We see signatures of both weak and strong sign-reversal. Increasing $\alpha$ beyond $\alpha_c$ results in strong sign-reversal, while change in the tilt parameter results in weak sign-reversal.
We fix the value of $\alpha_{12}$, and evaluate the fit parameters of $\sigma_{zz}(B)$ from Eq.~\ref{eq:sigma_2}. 
Fig.~\ref{fig:fournodes_lmc_colorplot_varyB5_vary_a14} (b) plots $\sigma_{zz}^{(2)}$ as a function of $B_5$ and $\alpha_{14}$. The contour $\alpha_{14c}$ where $\sigma_{zz}^{(2)}$ switches sign shows a dependence on $B_5$ as well. Therefore the contour $\alpha_{c}$ is in general a function of both $t_z$ and $B_5$. Fig.~\ref{fig:fournodes_lmc_colorplot_varyB5_vary_a14} (c) and (d) plot the parameters $B_0$ and $\sigma_{zz}^{(0)}$ obtained from Eq.~\ref{eq:sigma_2}, both of which display very interesting behavior as a result of varying $B_5$ and $\alpha_{14}$. In Fig.~\ref{fig:fournodes_lmc_colorplot_varyB5_vary_a14} (c), when $\alpha<\alpha_c(B_5)$, the sign of $B_0$ changes from negative to positive as $B_5$ changes sign from negative to positive. When $\alpha>\alpha_c(B_5)$, the change of sign is from positive to negative. At $\alpha=\alpha_c(B_5)$, there is strong sign-reversal resulting in sharp contrasting features on the both sides of $\alpha_c(B_5)$. On the other hand, in Fig.~\ref{fig:fournodes_lmc_colorplot_varyB5_vary_a14} (d), $\sigma_{zz}^{(0)}$ does not change sign as $B_5$ changes sign, but like $B_0$, it displays striking behavior around $\alpha_{c}(B_5)$ due to strong sign-reversal.  

Before closing this section, we also comment on the planar Hall effect in inversion asymmetric Weyl semimetals. Fig.~\ref{fig:fournodes_sxz_colorplot} (a) plots the planar Hall conductivity $\sigma_{xz}$ as a function of the angle $\gamma_5$ in the absence of an external magnetic field and presence of strain induced gauge field $B_5$. The PHC behaves as $\sim \sin(2\gamma_5)$ as in Fig.~\ref{fig:twonodes_tiltz_opp_same_phc_vs_gm5} (a). The contribution from the two time-reversed and opposite tilt Weyl node pairs adds up, while the contribution from two time-reversed and same tilt Weyl node pairs cancels out, and that is why we do not get a $\sim\sin(\gamma_5)$ trend as in Fig.~\ref{fig:twonodes_tiltz_opp_same_phc_vs_gm5} (b). In Fig.~\ref{fig:fournodes_sxz_colorplot} (b), we plot the change in the magnitude of the planar Hall conductivity upon infinitesimally increasing the intervalley strength $\alpha_{14}$. We again notice a region in the $B_5-t_z$ space where the variation of conductivity is anomalous, i.e. increasing intervalley scattering increases the magnitude of the conductivity. A similar plot is observed when we instead fix $\alpha_{14}$ and vary $\alpha_{12}$, therefore we do not explicitly plot this here. 

\section{Conclusions}
The sign of longitudinal magnetoconductivity in Weyl semimetals due to chiral anomaly has been a subject of intense research~\cite{spivak2016magnetotransport, das2019linear,imran2018berry,kim2014boltzmann, dantas2018magnetotransport,johansson2019chiral,das2019berry,cortijo2016linear,zyuzin2017magnetotransport,knoll2020negative,sharma2020sign,ahmad2021longitudinal,sharma2022revisiting, sharma2017chiral}. Almost unanimously, the sign of longitudinal magnetoconductivity has been agreed upon to be positive, at least in the limit of weak magnetic fields. However, various factors, such as tilting of the Weyl cones, strain and inhomogeneties in the material, qualitatively affect the LMC in Weyl semimetals. The interplay between various parameters, such as intervalley scattering, tilt, strain induced chiral gauge field, and the external magnetic field, leads to many striking features in both the longitudinal magnetoconductance and the planar Hall conductance of Weyl semimetals, which has been the focus of this work.  

In this work, we first show that the conventional method of assigning sign to magnetoconductivity, i.e., comparing the magnitude of conductivity for field $B$ with $B\pm\epsilon$ ($\epsilon$ being arbitrary),  leads to ambiguities when the system is subjected to strain. Specifically, the sign of magnetoconductivity could depend on the direction of the magnetic field. 
Thus there is a necessity to define weak sign-reversal and strong sign-reversal, both of which are qualitatively different, and result in qualitatively different responses. Weak sign-reversal, in general, leads to smooth changes in the fit parameters of the conductivity, while strong sign-reversal leads to very sharp changes. Weak sign-reversal is specifically is characterized by a change in the vertex and the axis of the parabola of conductivity with respect to the magnetic field, while strong sign-reversal is characterized by an opposite orientation, i.e., the direction in which the parabola opens is reversed.
Broadly speaking: (i) when strain induced chiral gauge field is absent and external magnetic field is present, strong intervalley scattering results in strong-sign reversal, (ii) when chiral gauge field is present and magnetic field is absent, the system, by default, shows strong sign-reversed state for both weak and strong intervalley scattering, (iii) when both chiral gauge and external magnetic field are present, there is both weak and strong sign-reversal. The latter is also experimentally the most relevant scenario, and we show that it leads to very striking phase plots that can be explored experimentally in current and upcoming experiments in Weyl semimetals. In practice, the parameters could be evaluated by fitting the conductivity from the experiments and that could give us insight into the strain, tilt, and dominant scattering mechanism in the system. We have also studied the effect of strain on the planar Hall conductance.
Another striking feature of anomalous variation of the planar Hall conductivity is also unraveled due to the rich interplay between the chiral gauge and external magnetic field, where the magnitude of conductivity can increase on increasing scattering strength.  

\textit{Acknowledgement:} AA acknowledges support from IIT Mandi HTRA. GS acknowledges support from grant SERB Grant No.
SRG/2020/000134. ST acknowledges support from grant  Grant
No. NSF 2014157. Discussions with Snehashish Nandy are gratefully acknowledged. 

\appendix
\section{Boltzmann formalism for magnetotransport}
Using the quasiclassical Boltzmann theory, we  study transport in Weyl semimetals in the limit of weak electric and magnetic fields. Since quasiclassical Boltzmann theory is valid away from the nodal point such that $\mu^2\gg \hbar v_F^2 e B$, therefore without any loss of generality we will assume that the chemical potential lies in the conduction band. The phenomenological Boltzmann equation for the non-equilibrium distribution function $f^\chi_\mathbf{k}$ can be expressed as~\cite{bruus2004many} 
\begin{align}
\left(\frac{\partial}{\partial t} + \dot{\mathbf{r}}^\chi\cdot \nabla_\mathbf{r}+\dot{\mathbf{k}}^\chi\cdot \nabla_\mathbf{k}\right)f^\chi_\mathbf{k} = \mathcal{I}_{\mathrm{coll}}[f^\chi_\mathbf{k}],
\label{Eq_boltz1}
\end{align}
where the collision term on the right-hand side of the equation incorporates the effects of scattering due to impurities.
In the presence of electric ($\mathbf{E}$) and magnetic ($\mathbf{B}$) fields, the semiclassical dynamics of the Bloch electrons is~\cite{son2012berry} 
\begin{align}
\dot{\mathbf{r}}^\chi &= \mathcal{D}^\chi \left( \frac{e}{\hbar}(\mathbf{E}\times \boldsymbol{\Omega}^\chi + \frac{e}{\hbar}(\mathbf{v}^\chi\cdot \boldsymbol{\Omega}^\chi) \mathbf{B} + \mathbf{v}_\mathbf{k}^\chi)\right) \nonumber\\
\dot{\mathbf{p}}^\chi &= -e \mathcal{D}^\chi \left( \mathbf{E} + \mathbf{v}_\mathbf{k}^\chi \times \mathbf{B} + \frac{e}{\hbar} (\mathbf{E}\cdot\mathbf{B}) \boldsymbol{\Omega}^\chi \right),
\end{align}
where $\mathbf{v}_\mathbf{k}^\chi$ is the band velocity, $\boldsymbol{\Omega}^\chi = -\chi \mathbf{k} /2k^3$ is the Berry curvature, and $\mathcal{D}^\chi = (1+e\mathbf{B}\cdot\boldsymbol{\Omega}^\chi/\hbar)^{-1}$. The self-rotation of Bloch wavepacket also gives rise to an orbital magnetic moment (OMM)~\cite{xiao2010berry} $\mathbf{m}^\chi_\mathbf{k}$. In the presence of magnetic field, the OMM shifts the energy dispersion as $\epsilon^{\chi}_{\mathbf{k}}\rightarrow \epsilon^{\chi}_{\mathbf{k}} - \mathbf{m}^\chi_\mathbf{k}\cdot \mathbf{B}$. Interestingly, the Berry curvature and the orbital magnetic moment turn out to be independent of the tilting of the Weyl cones.

The collision integral must take into account scattering between the two Weyl nodes (internode, $\chi\Longleftrightarrow\chi'$), as well as scattering withing a Weyl node (intranode, $\chi\Longleftrightarrow\chi$), and thus $\mathcal{I}_{\mathrm{coll}}[f^\chi_\mathbf{k}]$ can be expressed as 
\begin{align}
\mathcal{I}_{\mathrm{coll}}[f^\chi_\mathbf{k}] = \sum\limits_{\chi'}\sum\limits_{\mathbf{k}'} W^{\chi\chi'}_{\mathbf{k},\mathbf{k}'} (f^{\chi'}_{\mathbf{k}'} - f^\chi_\mathbf{k}),
\end{align}
where the scattering rate $W^{\chi\chi'}_{\mathbf{k},\mathbf{k}'}$ is given by~\cite{bruus2004many} 
\begin{align}
W^{\chi\chi'}_{\mathbf{k},\mathbf{k}'} = \frac{2\pi}{\hbar} \frac{n}{\mathcal{V}} |\langle \psi^{\chi'}_{\mathbf{k}'}|U^{\chi\chi'}_{\mathbf{k}\mathbf{k}'}|\psi^\chi_\mathbf{k}\rangle|^2 \delta(\epsilon^{\chi'}_{\mathbf{k}'}-\epsilon_F)
\label{Eq_W_1}
\end{align}
In the above expression $n$ is the impurity concentration, $\mathcal{V}$ is the system volume, $|\psi^\chi_\mathbf{k}\rangle$ is the Weyl spinor wavefunction (which is obtained by diagonalizing the low-energy Weyl Hamiltonian given in the main text), $U^{\chi\chi'}_{\mathbf{k}\mathbf{k}'}$ is the scattering potential, and $\epsilon_F$ is the Fermi energy. The scattering potential profile $U^{\chi\chi'}_{\mathbf{k}\mathbf{k}'}$ is determined by the nature of impurities. Here we restrict ourselves to only non-magnetic point-like impurity, but distinguish between intervalley and intravalley scattering.  This can be controlled independently in our formalism. Thus, the scattering matrix is momentum-independent but has a dependence on the chirality, i.e.,  $U^{\chi\chi'}_{\mathbf{k}\mathbf{k}'} = U^{\chi\chi'}\mathbb{I}$.

The distribution function is assumed to take the form $f^\chi_\mathbf{k} = f_0^\chi + g^\chi_\mathbf{k}$, where $f_0^\chi$ is the equilibrium Fermi-Dirac distribution function and $g^\chi_\mathbf{k}$ indicates the deviation from equilibrium. 
In the steady state, the Boltzmann equation (Eq.~\ref{Eq_boltz1}) takes the following form 
\begin{align}
&\left[\left(\frac{\partial f_0^\chi}{\partial \epsilon^\chi_\mathbf{k}}\right) \mathbf{E}\cdot \left(\mathbf{v}^\chi_\mathbf{k} + \frac{e\mathbf{B}}{\hbar} (\boldsymbol{\Omega}^\chi\cdot \mathbf{v}^\chi_\mathbf{k}) \right)\right]\nonumber\\
 &= -\frac{1}{e \mathcal{D}^\chi}\sum\limits_{\chi'}\sum\limits_{\mathbf{k}'} W^{\chi\chi'}_{\mathbf{k}\mathbf{k}'} (g^\chi_{\mathbf{k}'} - g^\chi_\mathbf{k})
 \label{Eq_boltz2}
\end{align}
The deviation $g^\chi_\mathbf{k}$ is assumed to be linearly proportional to the applied electric field 
\begin{align}
g^\chi_\mathbf{k} = e \left(-\frac{\partial f_0^\chi}{\partial \epsilon^\chi_\mathbf{k}}\right) \mathbf{E}\cdot \boldsymbol{\Lambda}^\chi_\mathbf{k}
\end{align}
We fix the direction of the applied external electric field to be along $+\hat{z}$, i.e., $\mathbf{E} = E\hat{z}$. Therefore only ${\Lambda}^{\chi z}_\mathbf{k}\equiv {\Lambda}^{\chi}_\mathbf{k}$, is relevant. Further, we rotate the magnetic field along the $xz$-plane such that it makes an angle $\gamma$ with respect to the $\hat{x}-$axis, i.e., $\mathbf{B} = B(\cos\gamma,0,\sin\gamma)$. When $\gamma=\pi/2$, the electric and magnetic fields are parallel to each other. Similarly, the strain induced chiral gauge field is rotated in the $xz$-plane, i.e,. $\mathbf{B_5}^\chi = \chi B_5(\cos\gamma_5,0,\sin\gamma_5)$. 
When $\gamma_5\neq \pi/2$, the electric and gauge field are non-collinear and this geometry will be useful in analyzing the strain induced planar Hall effect. Thus the net magnetic field at each valley becomes $\mathbf{B}^\chi\longrightarrow \mathbf{B}+\chi\mathbf{B_5}$.

Keeping terms only up to linear order in the electric field, Eq.~\ref{Eq_boltz2} takes the following form 
\begin{align}
\mathcal{D}^\chi \left[v^{\chi z}_{\mathbf{k}} + \frac{e B}{\hbar} \sin \gamma (\boldsymbol{\Omega}^\chi\cdot \mathbf{v}^\chi_\mathbf{k})\right] = \sum\limits_{\eta}\sum\limits_{\mathbf{k}'} W^{\eta\chi}_{\mathbf{k}\mathbf{k}'} (\Lambda^{\eta}_{\mathbf{k}'} - \Lambda^\chi_\mathbf{k})
\label{Eq_boltz3}
\end{align} 
In order to solve the above equation, we first define the valley scattering rate as follows
\begin{align}
\frac{1}{\tau^\chi_\mathbf{k}} = \mathcal{V} \sum\limits_{\eta} \int{\frac{d^3 \mathbf{k}'}{(2\pi)^3} (\mathcal{D}^\eta_{\mathbf{k}'})^{-1} W^{\eta\chi}_{\mathbf{k}\mathbf{k}'}}
\label{Eq_tau11}
\end{align}
Due to the tilting of the Weyl cones the azimuthal symmetry is destroyed even when the electric and magnetic fields are parallel to each other, and therefore all the integrations are performed over both $\theta$ and $\phi$. The radial integration is simplified due to the delta-function in Eq.~\ref{Eq_W_1}.

Substituting the scattering rate from Eq.~\ref{Eq_W_1} in the above equation, we have 
\begin{widetext}
\begin{align}
\frac{1}{\tau^\chi_\mathbf{k}} = \frac{\mathcal{V}N}{8\pi^2 \hbar} \sum\limits_{\eta} |U^{\chi\eta}|^2 \iiint{(k')^2 \sin \theta' \mathcal{G}^{\chi\eta}(\theta,\phi,\theta',\phi') \delta(\epsilon^{\eta}_{\mathbf{k}'}-\epsilon_F)(\mathcal{D}^\eta_{\mathbf{k}'})^{-1}dk'd\theta'd\phi'},
\label{Eq_tau1}
\end{align}
\end{widetext}
where $N$ now indicates the total number of impurities, and $ \mathcal{G}^{\chi\eta}(\theta,\phi,\theta',\phi') = (1+\chi\eta(\cos\theta \cos\theta' + \sin\theta\sin\theta' \cos(\phi-\phi')))$ is the Weyl chirality factor defined by the overlap of the wavefunctions. The Fermi wavevector contour $k^\chi$ is evaluated by equating the energy expression with the Fermi energy.
The three-dimensional integral in Eq.~\ref{Eq_tau1} is  reduced to just integration in $\phi'$ and $\theta'$. The scattering time ${\tau^\chi_\mathbf{k}}$ depends on the chemical potential ($\mu$), and is a function of the angular variables $\theta$ and $\phi$. 

\begin{align}
\frac{1}{\tau^\chi_\mu(\theta,\phi)} = \mathcal{V} \sum\limits_{\eta} \iint{\frac{\beta^{\chi\eta}(k')^3}{|\mathbf{v}^\eta_{\mathbf{k}'}\cdot \mathbf{k}'^\eta|}\sin\theta'\mathcal{G}^{\chi\eta}(\mathcal{D}^\eta_{\mathbf{k}'})^{-1} d\theta'd\phi'},
\label{Eq_tau2}
\end{align}
where $\beta^{\chi\eta} = N|U^{\chi\eta}|^2 / 4\pi^2 \hbar^2$. The Boltzmann equation (Eq~\ref{Eq_boltz3}) assumes the form  
\begin{align}
&h^\chi_\mu(\theta,\phi) + \frac{\Lambda^\chi_\mu(\theta,\phi)}{\tau^\chi_\mu(\theta,\phi)} =\nonumber\\ &\mathcal{V}\sum_\eta \iint {\frac{\beta^{\chi\eta}(k')^3}{|\mathbf{v}^\eta_{\mathbf{k}'}\cdot \mathbf{k}'^\eta|} \sin\theta'\mathcal{G}^{\chi\eta}(\mathcal{D}^\eta_{\mathbf{k}'})^{-1}\Lambda^\eta_{\mu}(\theta',\phi') d\theta'd\phi'}
\label{Eq_boltz4}
\end{align}
We make the following ansatz for $\Lambda^\chi_\mu(\theta,\phi)$
\begin{align}
\Lambda^\chi_\mu(\theta,\phi) &= (\lambda^\chi - h^\chi_\mu(\theta,\phi) + a^\chi \cos\theta +\nonumber\\
&b^\chi \sin\theta\cos\phi + c^\chi \sin\theta\sin\phi)\tau^\chi_\mu(\theta,\phi),
\label{Eq_Lambda_1}
\end{align}
where we solve for the eight unknowns ($\lambda^{\pm 1}, a^{\pm 1}, b^{\pm 1}, c^{\pm 1}$). The L.H.S in Eq.~\ref{Eq_boltz4} simplifies to $\lambda^\chi + a^\chi \cos\theta + b^\chi \sin\theta\cos\phi + c^\chi \sin\theta\sin\phi$. The R.H.S of Eq.~\ref{Eq_boltz4} simplifies to
\begin{align}
\mathcal{V}\sum_\eta \beta^{\chi\eta} \iint &f^{\eta} (\theta',\phi') \mathcal{G}^{\chi\eta} (\lambda^\eta - h^\eta_\mu(\theta',\phi') + a^\eta \cos\theta' +\nonumber\\
	&b^\eta \sin\theta'\cos\phi' + c^\eta \sin\theta'\sin\phi')d\theta'd\phi',
	\label{Eq_boltz5_rhs}
\end{align}
where the function
\begin{align}
f^{\eta} (\theta',\phi') = \frac{(k')^3}{|\mathbf{v}^\eta_{\mathbf{k}'}\cdot \mathbf{k}'^\eta|} \sin\theta' (\mathcal{D}^\eta_{\mathbf{k}'})^{-1} \tau^\chi_\mu(\theta',\phi')
\label{Eq_f_eta}
\end{align}
The above equations, when written down explicitly take the form of seven simultaneous equations to be solved for eight variables. The final constraint comes from the particle number conservation 
\begin{align}
\sum\limits_{\chi}\sum\limits_{\mathbf{k}} g^\chi_\mathbf{k} = 0
\label{Eq_sumgk}
\end{align}
Eq.~\ref{Eq_Lambda_1}, Eq.~\ref{Eq_boltz5_rhs}, Eq.~\ref{Eq_f_eta} and Eq.~\ref{Eq_sumgk} are solved together with Eq~\ref{Eq_tau2}, simultaneously for the eight unknowns ($\lambda^{\pm 1}, a^{\pm 1}, b^{\pm 1}, c^{\pm 1}$). Due to the complicated nature of the equations, all the two dimensional integrals w.r.t \{$\theta'$, $\phi'$\}, and the solution of the simultaneous equations are performed numerically. 

For the inversion asymmetric WSM with four Weyl nodes, the distribution function at each node can be represented by $f_\mathbf{k}^m$. Generalizing the formalism presented above, the collision integral must take into account scattering between multiple Weyl cones. Thus $\mathcal{I}_{\mathrm{coll}}[f^m_\mathbf{k}]$ can be expressed as 
\begin{align}
\mathcal{I}_{\mathrm{coll}}[f^m_\mathbf{k}] = \sum\limits_{p}\sum\limits_{\mathbf{k}'} W^{mp}_{\mathbf{k},\mathbf{k}'} (f^{p}_{\mathbf{k}'} - f^m_\mathbf{k}),
\end{align}
where $p$ runs over all the nodes, and scattering rate $W^{mp}_{\mathbf{k},\mathbf{k}'}$ is given by
\begin{align}
W^{mp}_{\mathbf{k},\mathbf{k}'} = \frac{2\pi}{\hbar} \frac{n}{\mathcal{V}} |\langle \psi^{p}_{\mathbf{k}'}|U^{mp}_{\mathbf{k}\mathbf{k}'}|\psi^m_\mathbf{k}\rangle|^2 \delta(\epsilon^{p}_{\mathbf{k}'}-\epsilon_F)
\end{align}
The scattering potential profile $U^{mp}_{\mathbf{k}\mathbf{k}'}$ can be chosen such that scattering between the nodes (internode) as well as within each node (intranode) is considered. Proceeding as before, we define ${\tau^m_\mathbf{k}}$ as

\begin{align}
\frac{1}{\tau^m_\mu(\theta,\phi)} = \mathcal{V} \sum\limits_{p} \iint{\frac{\beta^{mp}(k')^3}{|\mathbf{v}^p_{\mathbf{k}'}\cdot \mathbf{k}'^p|}\sin\theta'\mathcal{G}^{mp}(\mathcal{D}^p_{\mathbf{k}'})^{-1} d\theta'd\phi'},
\label{Eq_tau2m}
\end{align}
and the Boltzmann equation becomes
\begin{align}
&h^m_\mu(\theta,\phi) + \frac{\Lambda^m_\mu(\theta,\phi)}{\tau^m_\mu(\theta,\phi)} =\nonumber\\ &\mathcal{V}\sum_p \iint {\frac{\beta^{mp}(k')^3}{|\mathbf{v}^p_{\mathbf{k}'}\cdot \mathbf{k}'^p|} \sin\theta'\mathcal{G}^{mp}(\mathcal{D}^p_{\mathbf{k}'})^{-1}\Lambda^p_{\mu}(\theta',\phi') d\theta'd\phi'}.
\label{Eq_boltz4m}
\end{align}
Making the ansatz $\Lambda^m_\mu(\theta,\phi) = (\lambda^m - h^m_\mu(\theta,\phi) + a^m \cos\theta +b^m \sin\theta\cos\phi + c^m \sin\theta\sin\phi)\tau^m_\mu(\theta,\phi)$, and using the constraint for particle number conservation, the Boltzmann equation is reduced to a system of sixteen equations to be solved for sixteen unknowns.

\bibliography{biblio.bib}
\end{document}